\documentclass[journal,twoside,twocolumn,romanappendices]{IEEEtran-v17}
\newlength{\figwidth}
\usepackage{color}
\definecolor{links}{rgb}{0.7,0,0}   
\definecolor{urls}{rgb}{0,0,0.8}    
\definecolor{cites}{rgb}{0,0,0.8}   

\usepackage[nosort]{cite}        
\usepackage{url} 
\usepackage[intlimits]{amsmath}
\usepackage{bbm}
\usepackage{graphicx}
\usepackage[caption=false,font=footnotesize]{subfig}
\usepackage{paralist}
\usepackage{fancyref}
\usepackage[stretch=16,shrink=16,step=4]{microtype}
\usepackage{vmr-symbols-vecbold}
\usepackage{standard-macros}
\usepackage{steinmetz}
\usepackage{psfrag}
\usepackage{flushend}

\usepackage[colorlinks,hyperindex,linkcolor=links,citecolor=cites,urlcolor=urls]{hyperref} 

\makeatletter
\def\@IEEEinterspaceratioM{0.265}
\def\@IEEEinterspaceMINratioM{0.1651}
\def\@IEEEinterspaceMAXratioM{0.38}

\def\@IEEEinterspaceratioB{0.31}
\def\@IEEEinterspaceMINratioB{0.19}
\def\@IEEEinterspaceMAXratioB{0.38}
\@IEEEtunefonts
\makeatother
\let\Es\undefined
\newcommand{\Es}{\mathrm{E}}
\newcommand{\MR}{M}
\newcommand{\MT}{M}

\newcommand{\UL}{\text{ul}}
\newcommand{\DL}{\text{dl}}
\newcommand{\ULCOM}{\text{ul-clo}}
\newcommand{\ULSEP}{\text{ul-slo}}
\newcommand{\DLCOM}{\text{dl-clo}}
\newcommand{\DLSEP}{\text{dl-slo}}
\newcommand{\SISO}{\text{siso}}

\newcommand{\phxhato}{\tp{\vech}\bTheta_0 \hat{\vecx}_0}

\newcommand{\TRX}{\text{tx}}
\newcommand{\RX}{\text{rx}}
\begin{document}

\IEEEoverridecommandlockouts

\title{Capacity of SIMO and MISO Phase-Noise Channels with Common/Separate Oscillators}
%
%
\author{M. Reza Khanzadi,~\IEEEmembership{Student Member,~IEEE}, Giuseppe~Durisi,~\IEEEmembership{Senior Member,~IEEE,} Thomas Eriksson~\IEEEmembership{}
\thanks{This work was partly supported by the Swedish Foundation for Strategic Research under grant SM13-0028.}
\thanks{The material of this paper was presented in part at the 2015 IEEE International Conference on Communications, London, U.K.}
\thanks{M.R. Khanzadi is with the Department of Signals and Systems and the Department of Microtechnology and Nanoscience, Chalmers University of Technology, Gothenburg,
Sweden (e-mail: khanzadi@chalmers.se).}

\thanks{G. Durisi and T. Eriksson are with the Department of Signals and Systems, Chalmers University of Technology, Gothenburg,
Sweden (e-mail: \{durisi,thomase\}@chalmers.se).}}
%
%
%
\markboth{IEEE TRANSACTIONS ON COMMUNICATIONS}{Khanzadi~\MakeLowercase{\textit{et al.}}: Capacity of SIMO and MISO Phase-Noise Channels with Common/Separate Oscillators}
\maketitle

\begin{abstract}
In multiple antenna systems, phase noise due to instabilities of the radio-frequency (RF) oscillators, acts differently depending on whether the RF circuitries connected to each antenna are driven by separate (independent) local oscillators (SLO) or by a common local oscillator (CLO). In this paper, we investigate the high-SNR capacity of single-input multiple-output (SIMO) and multiple-output single-input (MISO) phase-noise channels for both the CLO and the SLO configurations.

Our results show that the first-order term in the high-SNR capacity expansion is the same for all scenarios (SIMO/MISO and SLO/CLO), and equal to $0.5\ln (\snr)$, where $\snr$ stands for the SNR. On the contrary, the second-order term, which we refer to as phase-noise number, turns out to be scenario-dependent. For the SIMO case, the SLO configuration provides a diversity gain, resulting in a larger phase-noise number than for the CLO configuration. For the case of Wiener phase noise, a diversity gain of at least $0.5 \ln(M)$ can be achieved, where $M$ is the number of receive antennas. For the MISO, the CLO configuration yields a higher phase-noise number than the SLO configuration. This is because with the CLO configuration one can obtain a coherent-combining gain through maximum ratio transmission (a.k.a. conjugate beamforming). This gain is unattainable with the SLO configuration.
\end{abstract}
\section{Introduction}
\label{sec:introduction}
%
%
%

Phase noise due to phase and frequency instability in the local local radio-frequency (RF) oscillators used in wireless communication links results in imperfect synchronization between transmitters and receivers, which degrades the system throughput~\cite{lapidoth02-10a,colavolpe05-09a,mehrpouyan12-09a,durisi14-03a}, especially when high-order modulation schemes are used to support high spectral efficiency.

A fundamental way to assess the impact of phase noise on the throughput of wireless links is to determine the corresponding Shannon capacity.
Unfortunately, a closed-form expression for the capacity of wireless channels impaired by phase noise is not available (although it is known that the capacity-achieving distribution has discrete amplitude and  uniform independent phase when the phase-noise process is stationary and memoryless, with uniform marginal distribution over $[0,2\pi)$~\cite{katz04-10a}).
Nevertheless, both asymptotic capacity characterizations for large signal-to-noise ratio (SNR) and nonasymptotic capacity bounds are available in the literature.
Specifically, Lapidoth~\cite{lapidoth02-10a} characterized the first two terms in the high-SNR expansion of the capacity of a single-input single-output (SISO) stationary phase-noise channel.
Focusing on memoryless phase-noise channels, Katz and Shamai~\cite{katz04-10a} provided upper and lower bounds on the capacity that are tight at high SNR. 
The results in~\cite{lapidoth02-10a,katz04-10a} have been generalized to block-memoryless phase-noise channels in~\cite{nuriyev05-03a,durisi12-08a}. Numerical methods for the calculation of the information rates achievable with specific modulation formats have been proposed in, e.g.,~\cite{dauwels08-a,barletta12-05a, barletta14-07a}.

In multiple-antenna systems, phase noise acts differently depending on whether the RF circuitries connected to each antenna are driven by \emph{separate} (independent) local oscillators (SLO) or by a \emph{common} local oscillator (CLO).
Although the CLO configuration is intuitively more appealing because it results in a single phase-noise process to be tracked, the SLO configuration is unavoidable when the spacing between antennas needed to exploit the available spatial degrees of freedom, and, hence, achieve multiplexing or diversity gains, is  large~\cite{gesbert02-12a,chizhik-02-01a}.
This occurs for example in multiple-antenna line-of-sight microwave backhaul links operating in the $20$--$40\GHz$ frequency band, where the  spacing between antennas required to exploit the available spatial degrees of freedom can be as large as few meters~\cite{durisi14-03a}.
In large-antenna-array systems \cite{marzetta10-11a,rusek13-01a,bjornson13-07a}, cost and packaging considerations may also make the SLO configuration attractive.

For the CLO configuration, a high-SNR capacity expansion together with finite-SNR capacity upper and lower bounds have been recently reported in~\cite{durisi13-02a,durisi14-03a}.
For both the CLO and the SLO configurations, the multiplexing gain   was partly characterized in~\cite{durisi13-06a}.
In \cite{pitarokoilis13-06a,bjornson13-07a,bjornson14circuit},  lower bounds on the sum-rate capacity for the case when multiple single-antenna users communicate with a base station equipped with a large antenna array (uplink channel) have been developed for both CLO and SLO. 
These bounds suggest that the SLO configuration yields a higher sum-rate capacity than the CLO configuration. 
However, it is unclear whether these lower bounds are tight.

\paragraph*{Contributions}
\label{ssec:contributions}
We consider the scenario where a multiple-antenna base station communicates with a single-antenna user over an AWGN channel impaired by phase noise and study
the first two terms in the high-SNR capacity expansion, for both the uplink (SIMO) and the downlink (MISO) channel, and for both CLO and SLO. 
We characterize the first term and provide bounds on the second term that are tight for some phase-noise models of practical interest.
Our findings are as follows.
The first-order term  in the high-SNR capacity expansion turns out to be the same in all four scenarios, and equal to $0.5\ln(\snr)$, where $\snr$ stands for the SNR. 
In contrast, the second-order term, which we denote as \emph{phase-noise number} (terminology borrowed from the fading literature~\cite{lapidoth03-10a}), 
takes different values in the four cases.
For the uplink channel,  the SLO phase-noise number is larger than the CLO one. 
Intuitively, this holds because the SLO configuration provides a diversity gain. 
For the specific case of Wiener phase noise~\cite{demir05-05a}, we show that a diversity gain of at least $0.5 \ln (\MR)$, where $\MR$ is the number of receive antennas, can be achieved.
This result provides a theoretical justification of the observation reported in~\cite{pitarokoilis13-06a,bjornson13-07a,bjornson14circuit} that SLO yields  a higher sum-rate capacity than CLO for the uplink channel.

For the downlink channel, the ordering is reversed: the CLO configuration results in a higher phase-noise number than the SLO configuration.
Coarsely speaking, this holds because CLO allows for maximum-ratio transmission (a.k.a. conjugate beamforming), which yields a coherent-combing gain, whereas this gain is lost in the SLO case.
For the case of Wiener phase noise, we determine numerically the extent to which the quality of the local oscillators in the SLO configuration must be improved to overcome the loss of coherent-combing gain.

Our results are derived under the assumption that the continuous-time phase noise process remains constant over the duration of the symbol time. This assumption allows us to obtain a discrete-time equivalent channel model by sampling at Nyquist rate.
As shown recently in \cite{ghozlan13-07a,ghozlan14-01a,barletta14-01a}, by dropping this assumption one may obtain drastically different high-SNR behaviors.
In the Wiener phase-noise case, for example, the first-order term in the high-SNR capacity expression was shown in~\cite{ghozlan13-07a} to  be at least as large as $0.5 \ln(\rho)$. 
However, it is unclear whether this lower bound is tight.  

\paragraph*{Notation}
\label{ssec:sotations}
%
Boldface letters such as $\veca$  and $\matA$ denote vectors and matrices, respectively. 
The operator $\diag(\cdot)$, applied to a vector $\veca$, generates a square diagonal matrix having the elements of $\veca$ on its main diagonal.
With $\normal(0,\sigma^2)$ and $\jpg(0,\sigma^2)$, we denote the probability distribution of a real Gaussian random variable and of a circularly symmetric complex Gaussian random variable with zero mean and variance $\sigma^2$.
Furthermore,~$\setU[0,2\pi)$ stands for the uniform distribution over the interval $[0,2\pi)$, and $\mathrm{Gamma}(\alpha,\beta)$ stands for the Gamma distribution with parameters $\alpha$ and $\beta$; specifically, if $s$ is $\mathrm{Gamma}(\alpha,\beta)$-distributed, its probability density function (pdf) $q_s(s)$ is
\begin{IEEEeqnarray}{rCL}
  q_s(s)=\frac{s^{\alpha-1}e^{-s/\beta}}{\beta^{\alpha}\Gamma(\alpha)}, \quad s\geq 0
\end{IEEEeqnarray}
where $\Gamma(\cdot)$ denotes the Gamma function.
Throughout the paper, all sums between angles (both random and deterministic) are performed modulus $2\pi$, although this is not always explicitly mentioned so as to keep the notation compact. 
For a given discrete-time vector-valued random process $\{\vectheta_k\}$, we denote the sequence $\{\vectheta_m,\dots,\vectheta_n\}$, $m<n$ as $\vectheta_m^n$. 
When $m=1$, we omit the subscript.
For two functions $f(\cdot)$ and $g(\cdot)$, the notation $f(x)=\landauo(g(x))$, $x\to\infty$, means that $\lim_{x\to\infty}\abs{f(x)/g(x)}=0$.
For a given complex vector $\vecb$, we  denote by $\phase{\vecb}$ the vector that contains the phase of the elements of $\vecb$.
Finally, $\ln(\cdot)$ denotes the natural logarithm.
\section{Review of the SISO Case}
\label{sec:SISO}
We start by reviewing the results obtained in~\cite{lapidoth02-10a} for the SISO case. The analysis of the uplink scenario in Section~\ref{sec:simo} and of the downlink scenario in Section~\ref{sec:miso} will rely on these results.

Consider the  discrete-time SISO phase-noise channel
\begin{IEEEeqnarray}{rCL}\label{eq:siso_io}
  y_k=e^{j\theta_k} h x_k+w_k, \quad k=1,\dots,n.
\end{IEEEeqnarray}
Here, $x_k$ denotes the input symbol at discrete-time instant~$k$. The constant~$h$ is the path-loss coefficient, which is assumed deterministic, time-invariant, and known to the transmitter and the receiver;~$\{w_k\}$ are the additive Gaussian noise samples, drawn independently from a~$\jpg(0,2)$ distribution.\footnote{As we shall see in, e.g., Appendix~\ref{sec:proof_simo_multiple_osc_memoryless_capacity}, normalizing the noise variance to $2$ will turn out convenient.} 
Finally, the phase-noise process~$\{\theta_{k}\}$ is assumed stationary, ergodic, independent of~$\{w_k\}$, and with finite differential-entropy rate\footnote{Note that the differential-entropy rate of the complex random process $\{e^{j\theta_k}\}$ is equal to $-\infty$. This means that the results obtained in~\cite{lapidoth03-10a} in the context of fading channel are not applicable to~\eqref{eq:siso_io}.}
\begin{IEEEeqnarray}{rCL}\label{eq:finite_entropy_rate}
  h(\{\theta_k\})>-\infty.
\end{IEEEeqnarray}
Under these assumptions, the capacity of the SISO phase-noise channel~\eqref{eq:siso_io} is given by
\begin{IEEEeqnarray}{rCL}\label{eq:memory_siso_capacity_definition}
  C(\snr)=\lim_{n \to \infty}\frac{1}{n}\sup I(y^n;x^n)
\end{IEEEeqnarray}
where the supremum is over all probability distributions on~$x^n=(x_1,\dots,x_n)$ that satisfy the average-power constraint 
\begin{IEEEeqnarray}{rCL}\label{eq:average_power_constraint}
  \frac{1}{n}\sumo{k}{n}\Ex{}{\abs{x_k}^2}\leq 2\rho.
\end{IEEEeqnarray} 
Here, $\snr$ can be thought of as the SNR (recall that we set the noise variance to~$2$; hence, the SNR is equal to half the signal power $2\snr$).
A closed-form expression for the capacity of the phase-noise channel is not available. 
Lapidoth~\cite{lapidoth02-10a} proved the following asymptotic characterization of~$C(\snr)$.
\begin{thm}[\cite{lapidoth02-10a}]
\label{thm:siso_capacity_Lapidoth}
The capacity of the SISO phase-noise channel~\eqref{eq:siso_io} is given by
\begin{IEEEeqnarray}{rCL}\label{eq:siso_capacity_expansion_form}
  C(\snr)&=& \eta\ln (\snr) + \chi + \landauo(1), \quad \snr\to\infty
\end{IEEEeqnarray} 
where $\eta=1/2$ and
\begin{IEEEeqnarray}{rCL}\label{eq:siso_phase_noise_number}
\chi=({1}/{2})\ln \lefto({\abs{h}^2}/{2}\right)+\ln( 2\pi)-h\lefto(\{\theta_k\}\right).
\end{IEEEeqnarray}
\end{thm}
%

The factor $\eta=1/2$ in~\eqref{eq:siso_capacity_expansion_form} is the so-called capacity \emph{prelog}, defined as the
asymptotic ratio between capacity and the logarithm of SNR as
SNR grows to infinity: $\eta = \lim_{\snr \to \infty}{C(\snr)}/{\ln (\snr)}$. 
The capacity prelog can be interpreted as the fraction of complex
dimensions available for communications in the limiting regime of high signal power, or equivalently vanishing noise variance~\cite{durisi11-08a}. 
For the phase-noise channel~\eqref{eq:siso_io}, only the amplitude $\abs{x_k}$ of the transmitted signal $x_k$ can be perfectly recovered in the absence of  additive noise, whereas the phase $\phase{x_k}$ is lost. Hence, the fraction of complex dimensions available for communication is~$\eta=1/2$.

We denote the second term in the high-SNR expansion \eqref{eq:siso_capacity_expansion_form} of~$C(\snr)$ as 
the~\emph{phase-noise number}~$\chi$
\begin{IEEEeqnarray}{rCL}\label{eq:phase_noise_number}
\chi=\lim_{\snr \to \infty}\left\{C(\snr)-\eta\ln (\snr) \right\}.
\end{IEEEeqnarray} 
%
We can see from \eqref{eq:siso_phase_noise_number} that the phase-noise number depends only on the statistics of the phase-noise process and on the path-loss coefficient~$h$. 
It is worth mentioning that the approximation $C(\snr)\approx \eta\ln (\snr) +\chi$, although based on a high-SNR capacity expansion, is often accurate already at low SNR values~\cite{katz04-10a,durisi12-08a,durisi14-03a}.
Next, we provide closed-form expressions for~$\chi$ for the phase-noise models commonly used in the wireless literature. 

\paragraph*{Noncoherent System}
Consider the case where the phase-noise process $\{\theta_k\}$ is  stationary and memoryless with uniform marginal distribution over $[0,2\pi)$. This scenario models accurately a noncoherent communication system where the phase of~$x_k$ is not used to transmit information~(see~\cite{katz04-10a}). 
The phase-noise number for this case can be readily obtained from~\eqref{eq:siso_phase_noise_number} by using that $h(\{\theta_k\})=\ln( 2\pi)$.

\begin{figure}[t]
\begin{center}
\psfrag{unwent}[][][0.8]{$0.5\ln(2\pi e \sigma^2_\Delta)$}
\psfrag{went}[][][0.8]{Eq.~\eqref{eq:wrapped_Gaussian_entropy}}
\psfrag{std}[][][0.8]{$\sigma_\Delta$~[degree]}
\psfrag{hd}[][][0.8]{Differential entropy~[bit]}
\includegraphics[width=2.6in]{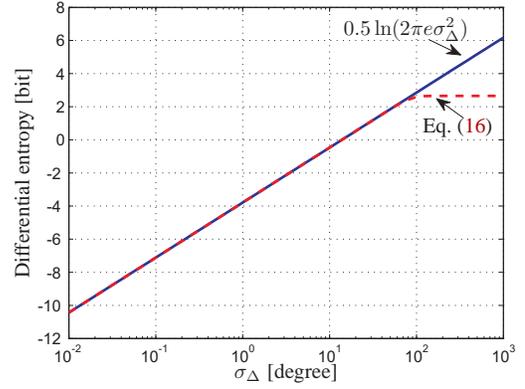}
\caption{Differential entropy of a wrapped and an unwrapped Gaussian random variable as a function of its standard deviation.}
\label{fig:Entropies_vs_STD}
\end{center}
\end{figure}

\paragraph*{Partially Coherent System}
When a phase tracker such as a phase-locked loop (PLL) is employed at the receiver, the output signal after phase tracking is impaired only by the residual phase error. 
Systems employing phase trackers are sometimes referred to as partially coherent~\cite{katz04-10a}. 
It is often accurate to assume that the residual phase-error process $\{\theta_k\}$ is stationary and memoryless.
Under this assumption, the phase-noise number for the partially-coherent case simplifies to 
\begin{IEEEeqnarray}{rCL}\label{eq:phase_noise_number_partially_coherent}
	\chi=({1}/{2})\ln \lefto({\abs{h}^2}/{2}\right)+\ln( 2\pi)-h\lefto(\theta\right)
\end{IEEEeqnarray}
where $\theta$ is the random variable modeling the residual phase error. 
When a PLL is used, the statistics of~$\theta$ are accurately described by a Tikhonov distribution 
\begin{IEEEeqnarray}{rCL}\label{eq:TikhonovDist}
	f_\theta(\theta)=\frac{e^{\lambda\cos\theta}}{2\pi I_0(\lambda)}
\end{IEEEeqnarray}
where $1/\lambda$ is the variance of~$\theta$, which depends on the oscillator quality and also on the parameters of the PLL \cite{Viterbi1963}. 
In this case, 
\begin{IEEEeqnarray}{rCL}\label{eq:diff_entropy_Tikhonov}
	h(\theta)&=&\ln\bigl(2\pi I_0(\lambda)\bigr)-\lambda {I_1(\lambda)}/{I_0(\lambda)}
\end{IEEEeqnarray}
where $I_0(\cdot)$ and $I_1(\cdot)$ stand for the modified Bessel functions of first kind and order $0$ and $1$, respectively.
\paragraph*{The Wiener Process}
The case of phase-noise process with memory is relevant when a free-running oscillator is used or when the phase tracker is not able to completely remove the memory of the phase-noise process~\cite{demir05-05a,Khanzadi2013_1}. The samples~$\{\theta_k\}$ of a free-running oscillator are typically modeled using a Wiener process~\cite{demir05-05a,colavolpe12-09a}, according to which
\begin{IEEEeqnarray}{rCL}\label{eq:Wiener_process}
	\theta_{k+1}=(\theta_k+\Delta_k)~{\text{mod}~(2\pi)}
\end{IEEEeqnarray}
where $\{\Delta_k\}$ are Gaussian random samples, independently drawn from a~$\normal (0,\sigma^2_\Delta)$ distribution. Hence, the sequence $\{\theta_k\}$ is a Markov process, i.e.,
\begin{IEEEeqnarray}{rCL}\label{eq:Markov_process}
	f_{\theta_k\given \theta_{k-1},\dots,\theta_{0}} = f_{\theta_k\given \theta_{k-1}}=f_{\Delta}
\end{IEEEeqnarray}
where the wrapped Gaussian distribution
\begin{IEEEeqnarray}{rCL}\label{eq:innovation_pdf}
	f_{\Delta}(\delta) = \suminf{l}\frac{1}{\sqrt{2\pi\sigma^2_\Delta}}\exp\lefto(-\frac{(\delta-2\pi l)^2}{2\sigma^2_\Delta}\right),\quad \delta\in[0,2\pi)\notag\\
\end{IEEEeqnarray}
is the pdf of the innovation $\Delta$ modulus $2\pi$. 
Under the assumption that the initial phase-noise sample~$\theta_0$ is uniformly distributed over~$[0,2\pi)$, the process~$\{\theta_k\}$ is stationary. Hence, its differential-entropy rate is given by the differential entropy of the innovation process
\begin{IEEEeqnarray}{rCL}\label{eq:Wiener_entropy_rate}
	h(\{\theta_k\})=h(\Delta).
\end{IEEEeqnarray}
The phase-noise number of the Wiener phase-noise channel can be readily obtained by using that the differential entropy of the wrapped Gaussian random variable $\Delta$ is given by~\cite{mardia09-a}
\begin{IEEEeqnarray}{rCL}\label{eq:wrapped_Gaussian_entropy}
	h(\Delta)= -\ln\lefto(\frac{\varphi(e^{-\sigma^2_\Delta})}{2\pi}\right)+2\sumo{n}{\infty}\frac{(-1)^n}{n}\frac{e^{-\sigma^2_\Delta(n^2+n)/2}}{1-e^{-n\sigma^2_\Delta}}\notag\\
\end{IEEEeqnarray}
where 
\begin{IEEEeqnarray}{rCL}
\varphi(e^{-\sigma^2_\Delta})=\prod_{l=1}^\infty\lefto(1-e^{-l\sigma^2_\Delta}\right). 
\end{IEEEeqnarray}
As shown in Fig.~\ref{fig:Entropies_vs_STD}, $h(\Delta)$ can be well-approximated by the differential entropy of an unwrapped $\normal (0,\sigma^2_\Delta)$ random variable
\begin{IEEEeqnarray}{rCL}\label{eq:wrapped_Gaussian_entropy_approx}
	h(\Delta)\approx ({1}/{2})\ln(2\pi e\sigma^2_\Delta)
\end{IEEEeqnarray}
whenever the standard deviation $\sigma_\Delta$ is below $55^{\circ}$. The oscillators commonly used in wireless transceivers result in a phase-noise standard variation that is well below $55^{\circ}$ \cite[Fig.~2]{khanzadi2014_C01a}. 

\section{Uplink Channel}
\label{sec:simo}
Building on the results reviewed in Section~\ref{sec:SISO}, we next analyze the uplink channel of a wireless communication system where a single-antenna terminal communicates with a base station equipped with $\MR$ antennas over an AWGN channel impaired by phase noise.
This yields the following $1\times \MR$ single-input multiple-output (SIMO) phase-noise channel:
\begin{IEEEeqnarray}{rCL}\label{eq:simo_io}
  \vecy_k=\bTheta_k\vech x_k+\vecw_k, \quad k=1,\dots,n.
\end{IEEEeqnarray}
Here, the matrix $\bTheta_k=\diag{([e^{j\theta_{1,k}},\dots,e^{j\theta_{\MR,k}}])}$ contains the phase-noise samples. 
We assume that, for each $m=1,\dots,\MR$, the phase-noise process $\{\theta_{m,k}\}$ is stationary, ergodic, independent of the additive-noise process $\{\vecw_k\}$, and has finite differential-entropy rate. 
Note that we do not necessarily assume that the phase-noise processes $\{\theta_{m,k}\}$,~$m=1,\dots,\MR$ are independent. 
It will turn out convenient to define also the phase-noise vector-valued process $\{\vectheta_k\}$ where $\vectheta_k=\tp{[\theta_{1,k},\dots,\theta_{\MR,k}]}$.
The vector $\vech=\tp{[h_1,\dots, h_{\MR}]}$ contains the path-loss coefficients, which, similarly to the SISO case, are assumed to be deterministic, time-invariant, and known to the transmitter and the receiver. Finally, the vector $\vecw_k=\tp{[w_{1,k},\dots, w_{\MR,k}]}$ contains the AWGN samples, which are drawn independently from a $\jpg(0,2)$ distribution.
Similarly to~\eqref{eq:memory_siso_capacity_definition}, the capacity of the SIMO phase-noise channel~\eqref{eq:simo_io}~is 
\begin{IEEEeqnarray}{rCL}\label{eq:memory_simo_capacity_definition}
  C(\snr)=\lim_{n \to \infty}\frac{1}{n}\sup I(\vecy^n;x^n)
\end{IEEEeqnarray}
where the supremum is over all probability distributions on~$x^n$ that satisfy the average-power constraint \eqref{eq:average_power_constraint}.

\subsection{Uplink, Common Local Oscillator (UL-CLO)}\label{sec:UL-CLO}
In the CLO configuration, we have that~$\theta_{1,k}=\dots=\theta_{\MR,k}=\theta_{k}$ for all $k$. Hence, the input-output relation~\eqref{eq:simo_io} simplifies to
\begin{IEEEeqnarray}{rCL}\label{eq:simo_common_osc_io}
  \vecy_k=e^{j\theta_k}\vech x_k+\vecw_k.
\end{IEEEeqnarray}
%
By projecting~$\vecy_k$ on~$\vech/\vecnorm{\vech}$, i.e., by performing coherent/maximal-ratio combining, we obtain a sufficient statistics for the detection of~$x_k$ from~$\vecy_k$.
 Through this projection, the SIMO phase-noise channel~\eqref{eq:simo_common_osc_io} is transformed into an equivalent SISO phase-noise channel with channel gain~$\vecnorm{\vech}$. 
Therefore, using Theorem~\ref{thm:siso_capacity_Lapidoth}, we conclude that the prelog for the UL-CLO case is~$\eta_\ULCOM=1/2$ and that the phase-noise number is
\begin{IEEEeqnarray}{rCL}\label{eq:PNN_simo_common_osc_memory}
  \chi_\ULCOM=({1}/{2})\ln \lefto({\vecnorm{\vech}^2}/{2}\right)+ \ln( 2\pi)-h(\{\theta_k\}).
\end{IEEEeqnarray}

\subsection{Uplink, Separate Local Oscillators (UL-SLO)}\label{sec:UL-SLO}
In the SLO case, the $\MR$ phase-noise processes~$\{\theta_{m,k}\}$, are independent and identically distributed (\iid) across the receive antennas. Hence, coherent combining does not yield a sufficient statistics. 
In Theorem~\ref{thm:PNN_simo_separate_osc_memory} below, we provide a characterization of the high-SNR capacity of $C(\snr)$, which holds irrespectively of the dependency between the $\MR$ phase-noise processes~$\{\theta_{m,k}\}$,~$m=1,\dots,\MR$.

\begin{thm}
\label{thm:PNN_simo_separate_osc_memory}
The prelog of the SIMO phase-noise channel \eqref{eq:simo_io} is given by $\eta_\UL={1}/{2}$.
Furthermore, the phase-noise number is bounded by
\begin{IEEEeqnarray}{rCL}\label{eq:PNN_simo_separate_osc_memory} 
\chi_\UL& \geq & ({1}/{2})\ln \lefto({\vecnorm{\vech}^2}/{2}\right) + 
\ln (2\pi)\notag\\&&-h\lefto(\phi_0\given \vectheta_0+\phi_0,\vectheta_{-\infty}^{-1}\right)  \IEEEyesnumber\IEEEyessubnumber\label{eq:PNN_simo_separate_osc_memory_LB} \\
\chi_\UL& \leq & ({1}/{2})\ln \lefto({\vecnorm{\vech}^2}/{2}\right) + 
\ln (2\pi)\notag\\&&-h\lefto(\phi_0\given \vectheta_0+\phi_0\right)+I\lefto(\vectheta_0;\vectheta_{-\infty}^{-1}\right) \IEEEyessubnumber  \label{eq:PNN_simo_separate_osc_memory_UB}
\end{IEEEeqnarray} 
where $\{\phi_k\}$ is a stationary memoryless process, with marginal distribution uniform over $[0,2\pi)$.
\end{thm}
\begin{IEEEproof}
See Appendix~\ref{sec:PNN_simo_separate_osc_memory}.
\end{IEEEproof}

\paragraph*{Remark~1}
A more accurate characterization of the phase-noise number may be obtained by adapting to the case of phase noise the tools developed in~\cite{lapidoth06-02a} for the analysis of stationary SIMO fading channels at high SNR.
We leave this refinement for future work.

\paragraph*{Remark~2}
The upper and lower bounds in~\eqref{eq:PNN_simo_separate_osc_memory} match when the phase noise processes are memoryless. 
Indeed, under this assumption,
\begin{IEEEeqnarray}{rCL}
\chi_\UL = ({1}/{2})\ln \lefto({\vecnorm{\vech}^2}/{2}\right) + 
\ln (2\pi)-h\lefto(\phi_0\given \vectheta_0+\phi_0\right).\IEEEeqnarraynumspace\label{eq:PNN_simo_separate_osc_memoryless} 
\end{IEEEeqnarray} 

\paragraph*{Remark~3} 
For the CLO case where~$\theta_{1,k}=\dots=\theta_{\MR,k}=\theta_{k}$ for all $k$, 
the bounds in \eqref{eq:PNN_simo_separate_osc_memory} match and reduce to \eqref{eq:PNN_simo_common_osc_memory}.
Indeed, for the lower bound we have that
\begin{IEEEeqnarray}{rCL}
\chi_\UL &\geq& \frac{1}{2}\ln \lefto(\frac{\vecnorm{\vech}^2}{2}\right) +
\ln (2\pi)-h\lefto(\phi_0\given \vectheta_0+\phi_0,\vectheta_{-\infty}^{-1}\right)\IEEEeqnarraynumspace\\
%
&=& \frac{1}{2}\ln \lefto(\frac{\vecnorm{\vech}^2}{2}\right) + 
\ln (2\pi)-h\lefto(\phi_0\given \theta_0+\phi_0,\theta_{-\infty}^{-1}\right)\\
%
%
&=& \frac{1}{2}\ln \lefto(\frac{\vecnorm{\vech}^2}{2}\right) + 
I\lefto(\phi_0; \theta_0+\phi_0\given\theta_{-\infty}^{-1}\right)\\
&=& \frac{1}{2}\ln \lefto(\frac{\vecnorm{\vech}^2}{2}\right) + 
h\lefto(\theta_0+\phi_0\given\theta_{-\infty}^{-1}\right)\notag\\&&-h\lefto(\theta_0+\phi_0\given\theta_{-\infty}^{-1},\phi_0\right)\\
&=& \frac{1}{2}\ln \lefto(\frac{\vecnorm{\vech}^2}{2}\right) + 
\ln (2\pi)-h\lefto(\theta_0 \given \theta_{-\infty}^{-1}\right)\label{eq:PNN_simo_COM_from__LB_SEP}=\chi_\ULCOM.
\end{IEEEeqnarray}
Here, in \eqref{eq:PNN_simo_COM_from__LB_SEP} we used that~$(\theta_0+\phi_0)\sim\setU[0,2\pi)$, which holds because $\phi_0\sim\setU[0,2\pi)$. 
For the upper bound, we have
\begin{IEEEeqnarray}{rCL}
\chi_\UL&\leq& \frac{1}{2}\ln \lefto(\frac{\vecnorm{\vech}^2}{2}\right) + 
\ln (2\pi)\notag\\&&-h\lefto(\phi_0\given \vectheta_0+\phi_0\right)+I\lefto(\vectheta_0;\vectheta_{-\infty}^{-1}\right)\\
&=& \frac{1}{2}\ln \lefto(\frac{\vecnorm{\vech}^2}{2}\right) + 
\ln (2\pi)\notag\\&&-h\lefto(\phi_0\given \theta_0+\phi_0\right)+I\lefto(\theta_0;\theta_{-\infty}^{-1}\right)\\
&=& \frac{1}{2}\ln \lefto(\frac{\vecnorm{\vech}^2}{2}\right) + 
I\lefto(\phi_0; \theta_0+\phi_0\right)+I\lefto(\theta_0;\theta_{-\infty}^{-1}\right)\\
&=& \frac{1}{2}\ln \lefto(\frac{\vecnorm{\vech}^2}{2}\right) + 
h\lefto(\theta_0+\phi_0\right)-h\lefto(\theta_0+\phi_0 \given \phi_0\right)\notag\\&&+h\lefto(\theta_0\right)-h\lefto(\theta_0\given\theta_{-\infty}^{-1}\right)\\
&=& \frac{1}{2}\ln \lefto(\frac{\vecnorm{\vech}^2}{2}\right) + 
\ln (2\pi)-h\lefto(\theta_0 \given \theta_{-\infty}^{-1}\right)\label{eq:PNN_simo_COM_from_SEP}=\chi_\ULCOM.\IEEEeqnarraynumspace
\end{IEEEeqnarray} 
\subsection{Discussion}
The fact that $\eta_\ULCOM=\eta_\ULSEP=\eta_\SISO$
%
comes perhaps as no surprise because adding multiple antennas at the receiver only (SIMO channel) does not yield spacial multiplexing gains. 
We next compare the phase-noise number of the CLO and the SLO configurations. 
We see from~\eqref{eq:PNN_simo_common_osc_memory} and~\eqref{eq:PNN_simo_separate_osc_memory} that the term $0.5\ln (\vecnorm{\vech}^2/{2})$ appears in the phase-noise number of both the CLO and SLO configuration. 
As already pointed out, in the CLO case this term comes from coherently combining the signals received at the $\MR$ antennas. 
Coherent combining is possible because, in the CLO case, the received signals at the different antennas are phase-shifted by the same random quantity. 
In the SLO case, however, coherent combining is not possible because the received signals at the different antennas are subject to independent random phase shifts. It turns out (see Appendix~\ref{sec:proof_simo_multiple_osc_memoryless_capacity} and Appendix~\ref{sec:proof_simo_multiple_osc_memoryless_arbitrary_capacity}) that a coherent-combining gain can be harvested regardless by separately decoding the amplitude and the phase of the transmitted signal, and by adding the square of the received signals when decoding the amplitude.

The CLO and SLO phase-noise numbers coincide in the noncoherent case (stationary, memoryless phase noise, with uniform marginal distribution over $[0,2\pi)$):
\begin{IEEEeqnarray}{rCL}
    \chi_\ULCOM=\chi_\ULSEP= ({1}/{2})\ln\lefto({\vecnorm{\vech}^2}/{2}\right).
\end{IEEEeqnarray} 
For phase-noise processes with memory, the CLO configuration results in a smaller phase-noise number than the SLO configuration. 
Indeed by rewriting the second and the third term on the RHS of~\eqref{eq:PNN_simo_common_osc_memory} as follows
\begin{IEEEeqnarray}{rCL}
  \ln(2\pi)-h(\{\theta_k\}) &=& I(\theta_0+\phi_0;\phi_0\given \theta_{-\infty}^{-1}) \\
  &=& \ln (2\pi) - h(\phi_0 \given \phi_0+\theta_0, \theta_{-\infty}^{-1})\IEEEeqnarraynumspace\label{eq:alt_CLO_UP_phase_noise_part}
\end{IEEEeqnarray}
where $\phi_0\sim\setU(0,2\pi]$ is independent of $\{\theta_k\}$, we see that the differential entropy on the RHS of~\eqref{eq:alt_CLO_UP_phase_noise_part} is larger than the differential entropy in the SLO phase-noise lower bound in~\eqref{eq:PNN_simo_separate_osc_memory_LB}.
To shed further light on the difference between the CLO and the SLO configuration, we now consider the special case of Wiener phase noise. 
For the CLO configuration, by substituting \eqref{eq:wrapped_Gaussian_entropy_approx} in \eqref{eq:Wiener_entropy_rate}, and then \eqref{eq:Wiener_entropy_rate} in \eqref{eq:PNN_simo_common_osc_memory} we obtain
\begin{IEEEeqnarray}{rCL}
  \chi_\ULCOM &\approx&({1}/{2})\ln \lefto({\vecnorm{\vech}^2}/{2}\right) + \ln (2\pi) - ({1}/{2}) \ln (2\pi e \sigma^2_\Delta)
.\IEEEeqnarraynumspace\label{eq:PNN_simo_common_osc_Wiener}
\end{IEEEeqnarray}

For the SLO configuration, we manipulate the lower-bound in \eqref{eq:PNN_simo_separate_osc_memory_LB} as follows:
\begin{IEEEeqnarray}{rCL}
\IEEEeqnarraymulticol{3}{l}{  \chi_\ULSEP
\geq \frac{1}{2}\ln \lefto(\frac{\vecnorm{\vech}^2}{2}\right) + 
\ln(2\pi)-h\lefto(\phi_0\given \vectheta_0+\phi_0,\vectheta_{-\infty}^{-1}\right)
}\IEEEeqnarraynumspace\label{eq:PNN_simo_separate_osc_Wiener_lower_bound_1}\\
&=& \frac{1}{2}\ln \lefto(\frac{\vecnorm{\vech}^2}{2}\right) + 
\ln(2\pi)-h\lefto(\phi_0\given \{\phi_0+\Delta_{m,-1}\}_{m=1}^{\MR}\right)\label{eq:PNN_simo_separate_osc_Wiener_lower_bound_2}\\
&=& ({1}/{2})\ln \lefto({\vecnorm{\vech}^2}/{2}\right) + 
\ln(2\pi)\notag\\&&-h\lefto(\phi_0\Big | \{\phi_0+\Delta_{m,-1}\}_{m=1}^{\MR},\phi_0+\frac{1}{\MR}\sumo{m}{\MR}\Delta_{m,-1}\right)\label{eq:PNN_simo_separate_osc_Wiener_lower_bound_3}\\
&\geq& ({1}/{2})\ln \lefto({\vecnorm{\vech}^2}/{2}\right) + 
\ln(2\pi)\notag\\&&-h\lefto(\phi_0\Big |\phi_0+\frac{1}{\MR}\sumo{m}{\MR}\Delta_{m,-1}\right)\label{eq:PNN_simo_separate_osc_Wiener_lower_bound_4}\\
&=& \frac{1}{2}\ln \lefto(\frac{\vecnorm{\vech}^2}{2}\right) + 
\ln(2\pi)-h\lefto(\frac{1}{\MR}\sumo{m}{\MR}\Delta_{m,-1}\right)\\
&\approx& ({1}/{2})\ln \lefto({\vecnorm{\vech}^2}/{2}\right) + 
\ln(2\pi)-({1}/{2}) \ln\lefto(2\pi e {\sigma^2_\Delta}/{\MR}\right).\label{eq:PNN_simo_separate_osc_Wiener_final}
\end{IEEEeqnarray}
Here, \eqref{eq:PNN_simo_separate_osc_Wiener_lower_bound_2} follows by \eqref{eq:Wiener_entropy_rate} and by the Markov property of the Wiener process; \eqref{eq:PNN_simo_separate_osc_Wiener_lower_bound_3} holds because~$\phi_0$ and  $\phi_0+\sumo{m}{\MR}\Delta_{m,-1}/\MR$ are conditionally independent given $\{\phi_0+\Delta_{m,-1}\}_{m=1}^{\MR}$; in \eqref{eq:PNN_simo_separate_osc_Wiener_lower_bound_4} we used again that conditioning does not increase differential entropy; finally, \eqref{eq:PNN_simo_separate_osc_Wiener_final} follows from \eqref{eq:wrapped_Gaussian_entropy_approx}.

By comparing~\eqref{eq:PNN_simo_common_osc_Wiener} and~\eqref{eq:PNN_simo_separate_osc_Wiener_final} we see that for the Wiener  phase-noise case,~$\chi_\ULSEP \geq \chi_\ULCOM$. 
This gain can be explained as follows: in the SLO case, we have $\MR$ independent noisy observations of the phase of the transmitted signal. 
These independent noisy observations can be used to improve the estimation of the transmitted phase. 
Specifically, a diversity gain at least equal to~$0.5 \ln (\MR)$ can be achieved by using separate oscillators instead of a common oscillator. 
Equivalently, in order to obtain equal phase-noise numbers in the CLO and SLO configurations, the phase-noise variance~$\sigma^2_\Delta$ in the CLO case must be at least~$\MR$ times lower than in the SLO case. 
For large-antenna arrays, we expect the throughput gains resulting from the SLO configuration to occur only at very high SNR. Indeed, whereas in the CLO case 
the phase-noise tracker can leverage on the (large) antenna array gain, which yields a fast convergence to the high-SNR asymptotics, this is not the case in the SLO configuration, where each phase-noise process needs to be tracked separately, without relying on any array gain.  
It is perhaps also worth mentioning that the SLO gains cannot be achieved in the CLO case simply by independently phase-shifting the signal received at each antenna. In fact, this strategy does not even achieve the CLO phase-noise number~\eqref{eq:PNN_simo_common_osc_memory}.

A configuration that is perhaps more relevant from a practical point of view is the one where the $M$ phase-noise processes $\{\theta_{m,k}\},~m=1,\dots,M$ result from the sum of the phase-noise contribution $\theta_k^{(\TRX)}$ at the transmitter and of $M$ independent phase-noise contributions $\{\theta_{m,k}^{(\RX)}\},~m=1,\dots,M$ at the receivers.
Assuming that both $\theta_k^{(\TRX)}$ and $\{\theta_{m,k}^{(\RX)}\}$ evolve according to independent Wiener processes with iid innovations $\Delta_{k}^{(\TRX)}\sim \normal(0,\sigma^2_{\Delta,\TRX})$ and $\Delta_{m,k}^{(\RX)}\sim \normal(0,\sigma^2_{\Delta,\RX})$, we obtain 
\begin{IEEEeqnarray}{rCL}
 \chi_\ULSEP &\geq& ({1}/{2})\ln \lefto({\vecnorm{\vech}^2}/{2}\right) + 
\ln(2\pi)\notag\\
&&-h\lefto(\Delta_{-1}^{(\TRX)} +\frac{1}{\MR}\sumo{m}{\MR}\Delta_{m,-1}^{(\RX)}\right)\\
&\approx& ({1}/{2})\ln \lefto({\vecnorm{\vech}^2}/{2}\right) + 
\ln(2\pi)\notag\\
&&-({1}/{2}) \ln\lefto(2\pi e \left( \sigma^2_{\Delta,\TRX}+ {\sigma^2_{\Delta,\RX}}/{\MR}\right)\right).\label{eq:PNN_simo_separate_indep_osc_Wiener_final}
\end{IEEEeqnarray}
Here, \eqref{eq:PNN_simo_separate_indep_osc_Wiener_final} follows by proceeding as in \eqref{eq:PNN_simo_separate_osc_Wiener_lower_bound_1}--\eqref{eq:PNN_simo_separate_osc_Wiener_final}. The case where a single oscillator is used at the receiver can be obtained from \eqref{eq:PNN_simo_separate_indep_osc_Wiener_final} by setting $M=1$. Also for this setup, we see that using independent oscillators at the receiver is advantageous, although the gain is smaller than what suggested by \eqref{eq:PNN_simo_separate_osc_Wiener_final}.
\section{Downlink Channel}
\label{sec:miso}
We next analyze the downlink channel, i.e., the scenario where a base station equipped with~$\MT$ antennas communicates with a single-antenna terminal. This yields the following~$\MT \times 1$ multiple-input single-output (MISO) phase-noise channel
\begin{IEEEeqnarray}{rCL}\label{eq:miso_io}
  y_k&=& \tp{\vech}\bTheta_k \vecx_k+w_k.
\end{IEEEeqnarray}
Here, the phase-noise process $\{\bTheta_k\}$ and the path-loss vector $\vech$ are defined as in Section~\ref{sec:simo};~$\vecx_k=\tp{[x_{1,k},\dots,x_{\MT,k}]}$, where~$x_{m,k}$ denotes the  symbol  transmitted from antenna $m$ at time instant $k$; finally,~$\{w_k\}$ is the additive noise process, with samples drawn independently from a~$\jpg(0,2)$ distribution. 
Similarly to~\eqref{eq:memory_siso_capacity_definition} and~\eqref{eq:memory_simo_capacity_definition}, the capacity of the MISO phase-noise channel~\eqref{eq:miso_io} is
\begin{IEEEeqnarray}{rCL}\label{eq:miso_memory_capacity_def}
  C(\snr)=\lim_{n \to \infty}\frac{1}{n}\sup I(y^n;\vecx^n)
\end{IEEEeqnarray}
where the supremum is over all probability distributions on  $\vecx^n=(\vecx_1,\dots,\vecx_n)$ that satisfy the average-power constraint 
\begin{IEEEeqnarray}{rCL}\label{eq:miso_power_constraint}
  \frac{1}{n}\sumo{k}{n}\Ex{}{\vecnorm{\vecx_k}^2}\leq 2\rho.
\end{IEEEeqnarray} 

\subsection{Downlink, Common Local Oscillator (DL-CLO)}\label{sec:DL-CLO}
In the CLO case, we have that~$\theta_{1,k}=\dots=\theta_{\MT,k}=\theta_k$ for all~$k$. 
Hence, the input-output relation~\eqref{eq:miso_io} simplifies to 
\begin{IEEEeqnarray}{rCL}\label{eq:miso_common_osc_io}
	  y_k&=& e^{j\theta_k}\tp{\vech} \vecx_k+w_k.
\end{IEEEeqnarray}
Maximum ratio transmission, i.e., setting~$\vecx_k=s_k\conj{\vech} /\vecnorm{\vech}$, with $\{s_k\}$ chosen so that~\eqref{eq:miso_power_constraint} holds is capacity achieving. 
Indeed, set $s_k=\tp{\vech}\vecx_k/\vecnorm{\vech}$. 
Then
\begin{align}
  I(y_k;\vecx_k)&= I(y_k;\vecx_k,s_k) - I(y_k;s_k\given\vecx_k)\\&=I(y_k; s_k) +\underbrace{I(y_k;\vecx_k \given s_k)}_{=0} - I(y_k;s_k\given\vecx_k) \\
  &\leq I(y_k;s_k).
\end{align}
Here, the first equality follows from the chain rule for mutual information, the second equality follows because $y_k$ and $\vecx_k$ are conditionally independent given $s_k$ (see~\eqref{eq:miso_common_osc_io}), and in the last step we used that mutual information is nonnegative.
Note now that the upper bound is tight whenever $I(y_k;s_k\given\vecx_k)=0$. This is achieved by choosing $\vecx_k$ so that the transformation $\vecx_k\mapsto s_k$ is invertible.
This implies that, in order to achieve capacity, one should set $\vecx_k=s_k \conj{\vech}/\vecnorm{\vech}$.

With conjugate beamforming,  the MISO channel is transformed into a SISO channel.
Hence, Theorem~\ref{thm:siso_capacity_Lapidoth} allows us to conclude that, for the DL-CLO case, $\eta_\DLCOM={1}/{2}$ and
\begin{IEEEeqnarray}{rCL}
	\chi_\DLCOM&=&({1}/{2})\ln \lefto({\vecnorm{\vech}^2}/{2}\right)+ \ln(2\pi)-h(\{\theta_k\})\label{eq:PNN_miso_common_osc_memory}.
\end{IEEEeqnarray}
%

%

\subsection{Downlink, Separate Local Oscillator (DL-SLO)}\label{sec:DL-SLO}
In Theorem~\ref{thm:miso_multiple_osc_memory_upper_bound} below, we characterize the prelog and provide bounds on the phase-noise number of the  MISO phase-noise channel \eqref{eq:miso_io}. 
Afterwards, we shall discuss specific phase-noise models for which the bounds on the phase-noise number are tight.
Note that Theorem~\ref{thm:miso_multiple_osc_memory_upper_bound} holds irrespectively of the dependency between the $\MR$ phase-noise processes~$\{\theta_{m,k}\}$.
\begin{thm}
\label{thm:miso_multiple_osc_memory_upper_bound}
The prelog of the MISO phase-noise channel~\eqref{eq:miso_io} is given by $\eta_\DL = {1}/{2}$.
Furthermore, the phase-noise number is bounded by
\begin{IEEEeqnarray}{rCL}\label{eq:miso_PNN_upper_bound}
\chi_\DL &\geq& \ln (2\pi)\notag\\&&+\max_{m=1,\dots,\MT} \Biggr\{ \frac{1}{2}\ln \lefto(\frac{\abs{h_m}^2}{2}\right)-h\lefto(\{\theta_{m,k}\}\right)\Biggr\}\IEEEyesnumber\IEEEyessubnumber\IEEEeqnarraynumspace\label{eq:miso_PNN_upper_bound_LB}\\ 
\chi_\DL &\leq& \ln (2\pi)+\underset{\vecnorm{\hat{\vecx}}=1}{\sup }\Biggl\{\frac{1}{2}\ln\lefto(\frac{1}{2}\Ex{}{\abs{\tp{\vech}\bTheta_0 \hat{\vecx}}^2}\right)\Biggr\}\notag\\&&-\underset{\vecnorm{\hat{\vecx}}=1}{\inf}\Bigl\{h\bigl(\phase{\tp{\vech}\bTheta_0 \hat{\vecx}} \given \vectheta_{-\infty}^{-1}\bigr)\Bigr\}\IEEEyessubnumber\label{eq:miso_PNN_upper_bound_UB}
\end{IEEEeqnarray}
where $\hat{\vecx}$ is a unit-norm vector in $\complexset^{\MT}$.
\end{thm}
\begin{IEEEproof}
See Appendix~\ref{sec:proof_miso_multiple_osc_memory_upper_bound}.
\end{IEEEproof}

\paragraph*{Remark~4}
The lower bound on the phase-noise number in~\eqref{eq:miso_PNN_upper_bound_LB} is achieved by antenna selection, i.e., by activating only the transmit antenna that leads to the largest SISO phase-noise number. The other~$\MT-1$ transmit antennas are switched off. 
\paragraph*{Remark~5}
The bounds on the phase-noise number reported in~\eqref{eq:miso_PNN_upper_bound} may  be tightened using the tools developed in~\cite{moser06-07a,moser09-06a} in the context of MIMO fading channels. 
This tightening is left for future work.

\paragraph*{Remark~6}
In the CLO case where~$\theta_{1,k}=\dots=\theta_{1,k}=\theta_k$ for all~$k$, the upper bound in \eqref{eq:miso_PNN_upper_bound_UB} is tight.
Indeed,
\begin{IEEEeqnarray}{rCL}
    \chi_\DL&\leq& \ln(2\pi)+\underset{\vecnorm{\hat{\vecx}}=1}{\sup }\Biggl\{\frac{1}{2}\ln\lefto(\frac{1}{2}\Ex{}{\abs{\tp{\vech}\bTheta_0 \hat{\vecx}}^2}\right)\Biggr\}\notag\\&&-\underset{\vecnorm{\hat{\vecx}}=1}{\inf }\Bigl\{h\lefto(\phase{\tp{\vech}\bTheta_0 \hat{\vecx}} \given \vectheta_{-\infty}^{-1}\right)\Bigr\}\\
&=& \ln(2\pi)+\underset{\vecnorm{\hat{\vecx}}=1}{\sup }\Bigg\{\frac{1}{2}\ln\lefto(\frac{1}{2}\abs{\tp{\vech} \hat{\vecx}}^2\right)\Biggr\}\notag\\&&
-\underset{\vecnorm{\hat{\vecx}}=1}{\inf}\bigl\{ h\lefto(\phase{\tp{\vech} \hat{\vecx}}+\theta_0 \given \theta_{-\infty}^{-1}\right)\bigr\}\\
%
%
%
&=&\ln(2\pi)+({1}/{2})\ln \lefto({\vecnorm{\vech}^2}/{2}\right)-h(\{\theta_k\})\label{eq:miso_upper_bound_common_from_general_2}
=\chi_\DLCOM.\IEEEeqnarraynumspace
\end{IEEEeqnarray}
%
%
Note that the lower bound in~\eqref{eq:miso_PNN_upper_bound_LB} is not tight because antenna selection is not optimal for the CLO case.
The upper and lower bounds in~\eqref{eq:miso_PNN_upper_bound} match
when the phase-noise processes are independent across antennas, and have uniform marginal distributions over~$[0,2\pi)$. 
This occurs in noncoherent systems and for the Wiener model. 
We formalize this result below.
\begin{thm}\label{thm:miso_multiple_osc_memory_uniform_marginal}
The phase-noise number of the MISO phase-noise channel~\eqref{eq:miso_io} under the additional assumptions that the $M$ phase-noise processes $\{\theta_{m,k}\}$, $m=1,\dots,M$ 
i)~are independent and identically distributed (i.i.d.) across antennas (SLO configuration), 
ii)~have uniform marginal distributions over $[0,2\pi)$, is given by
\begin{IEEEeqnarray}{rCL}\label{eq:miso_multiple_osc_memory_uniform_marginal}
  \chi_\DLSEP&=&\frac{1}{2}\max_{m=1,\dots,\MT}\ln \lefto(\frac{\abs{h_m}^2}{2}\right)+\ln(2\pi)-h\lefto(\{\theta_k\}\right) \IEEEeqnarraynumspace
\end{IEEEeqnarray}
where $h\lefto(\{\theta_k\}\right)$ is the differential-entropy rate of one of the i.i.d. phase-noise processes.
\end{thm}
\begin{IEEEproof}
See Appendix~\ref{sec:proof_miso_multiple_osc_memory_uniform_marginal}.
\end{IEEEproof}  

We next compare $\chi_\DLCOM$ and $\chi_\DLSEP$. For the noncoherent case, we have that
\begin{IEEEeqnarray}{rCL}
  &&\chi_\DLCOM=({1}/{2})\ln \lefto({\vecnorm{\vech}^2}/{2}\right)\label{eq:miso_multiple_COM_PNN_noncoherent}\\
  &&\chi_\DLSEP=({1}/{2})\max_{m=1,\dots,\MT}\ln \lefto({\abs{h_m}^2}/{2}\right)\label{eq:miso_multiple_SEP_PNN_noncoherent}.
\end{IEEEeqnarray}
For the Wiener case, by substituting \eqref{eq:wrapped_Gaussian_entropy_approx} in \eqref{eq:Wiener_entropy_rate} and then \eqref{eq:Wiener_entropy_rate} in \eqref{eq:PNN_miso_common_osc_memory} and in~\eqref{eq:miso_multiple_osc_memory_uniform_marginal}, we obtain
\begin{IEEEeqnarray}{rCL}
  \chi_\DLCOM&\approx&({1}/{2})\ln \lefto({\vecnorm{\vech}^2}/{2}\right)+\ln(2\pi)-({1}/{2})\ln(2\pi e\sigma^2_\Delta)\label{eq:miso_multiple_COM_PNN_Wiener}\IEEEeqnarraynumspace\\
  \chi_\DLSEP&\approx&({1}/{2})\max_{m=1,\dots,\MT}\ln \lefto({\abs{h_m}^2}/{2}\right)+\ln(2\pi)\notag\\&&-({1}/{2})\ln(2\pi e\sigma^2_\Delta)\label{eq:miso_multiple_SEP_PNN_Wiener}.
\end{IEEEeqnarray}
In both the noncoherent and the Wiener case, we see that the SLO configuration results in no coherent-combining gain: $\vecnorm{\vech}^2$ is replaced by $\underset{m=1,\dots,\MT}{\max}\abs{h_m}^2$.
The resulting throughput loss is most pronounced when the entries of $\vech$ have all the same magnitude.
To shed further light on this loss, we depart from the model we considered so far, where the $\{h_m\}$,~$m=1,\dots,\MT$ are deterministic, and move to a quasi-static fading model \cite[p.~2631]{biglieri98-10a},\cite[Sec.~5.4.1]{tse05a}, where the~$\{h_m\}$ are independently drawn from a $\jpg(0,1)$ distribution and stay constant over the duration of a codeword. We also assume that the~$\{h_m\}$ are perfectly known to the transmitter and the receiver. 
In this scenario, $0.5\ln(\snr)+\chi$, where $\chi$ is now a function of the instantaneous channel gains, is the rate supported by the channel in the high-SNR regime, for a given channel realization. 

\begin{figure*}[!tb]
        \centering
         \subfloat[]{
\psfrag{COM}[][][0.7]{CLO}
\psfrag{SEP}[][][0.7]{SLO}
\psfrag{sd6}[][][0.7]{$\sigma_\Delta=6^\circ$}
\psfrag{sd1}[][][0.7]{$\sigma_\Delta=2.34^\circ$}
\psfrag{xLabel}[][][0.7]{$R$~[bit/channel use]}
\psfrag{yLabel}[][][0.7]{$\varepsilon=\text{Pr}\{0.5\ln(\snr)+\chi\leq R\}$}  \includegraphics[width=2.6in]{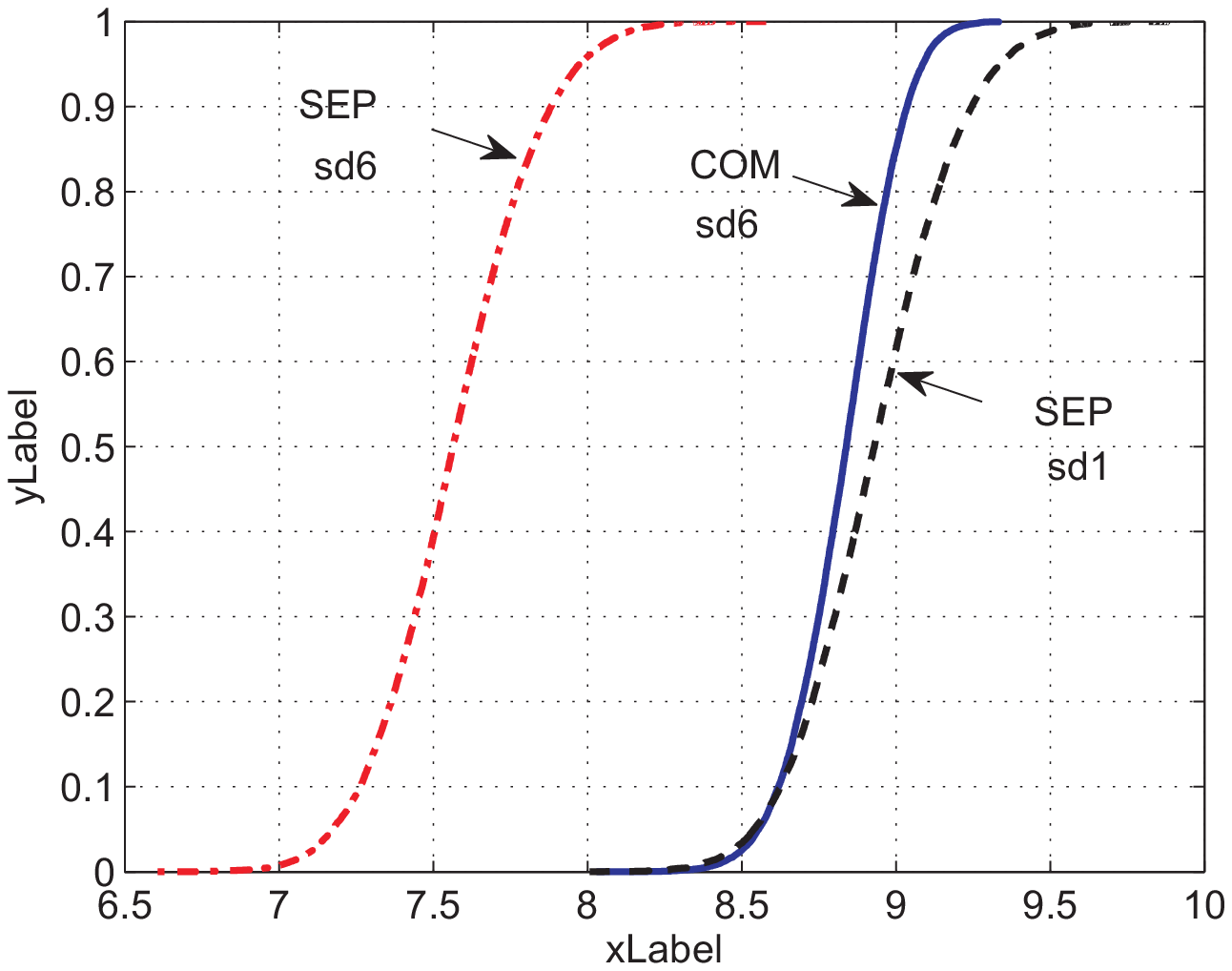}
          \label{fig:OutageProbability_MISO_Wiener_20MT}
          }
      \hspace{0.3cm}
      \subfloat[]{
\psfrag{xlabel}[][][0.7]{Number of antennas~$M$}
\psfrag{ylabel}[][][0.7]{$\Delta R$~[bit/channel use]}               \includegraphics[width=2.6in]{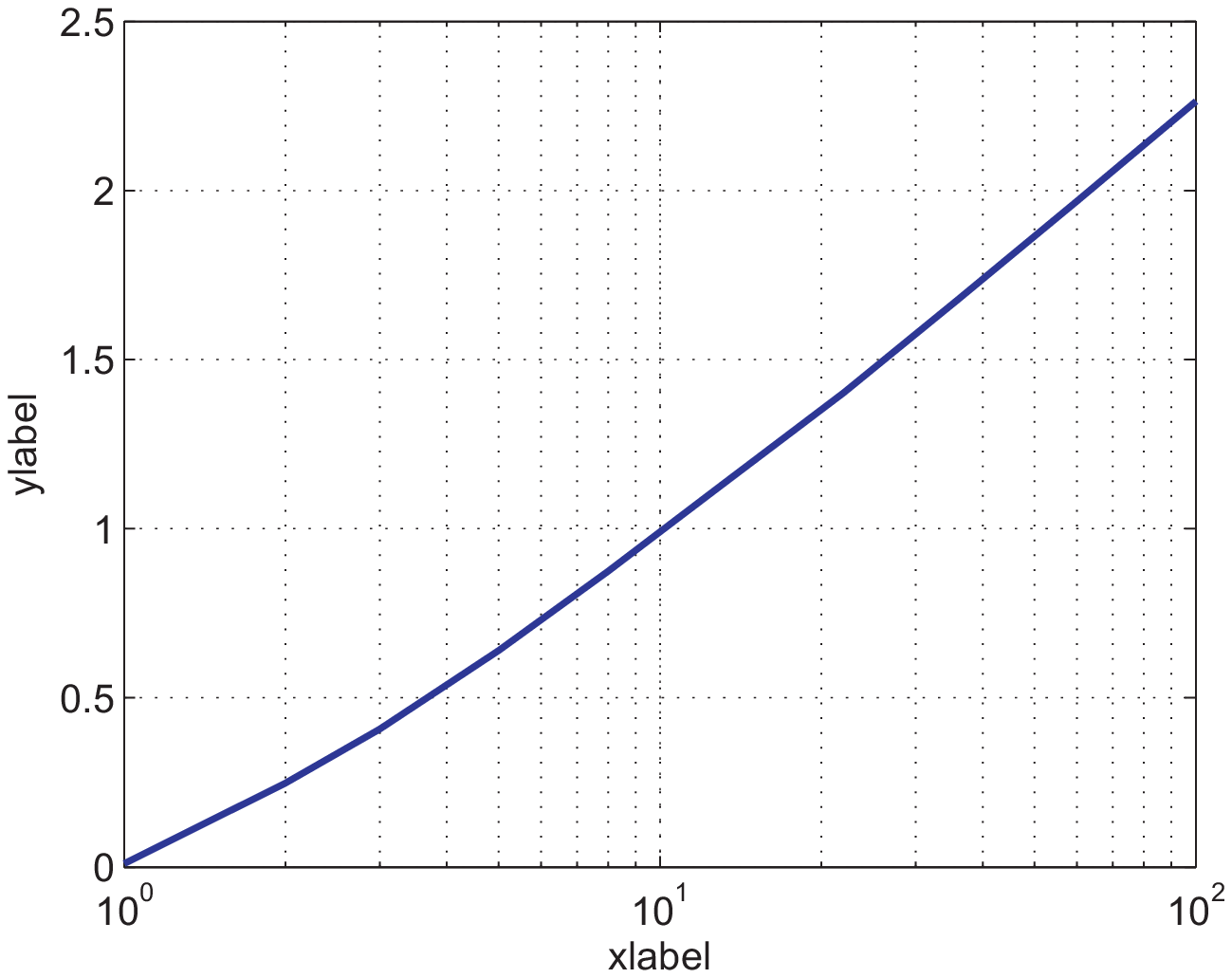}
                \label{fig:Diff_OutageProb_vsM_var6_log}
                }
        \caption{(a)~High-SNR approximation of the outage probability for the CLO and the SLO configurations. A Wiener phase-noise model is considered. Furthermore, $\MT=20$, and~$\rho=20$~dB. (b)~Difference $\Delta R$ between the outage capacity of the CLO and the SLO configurations for a target outage probability of $\varepsilon=0.1$, as a function of the number of antennas.}
\label{fig:MSEs}
\end{figure*}

In Fig.~\ref{fig:OutageProbability_MISO_Wiener_20MT}, we plot the cumulative distribution function of $0.5\ln(\snr)+\chi$, which is a high-SNR approximation of the outage capacity. 
We consider the case of Wiener phase noise with standard deviation $\sigma_\Delta=6^\circ$ and set~$\MT=20$ and~$\rho=20$~dB. 
For a given outage probability, the rate supported in the SLO case is smaller than that in the CLO case. 
For example, for a target outage probability of $\varepsilon=0.1$, the  rate supported in the SLO case is $1.36$~bit/channel use lower than that in the CLO case. 
To achieve the same rate at $\varepsilon=0.1$, the standard deviation $\sigma_{\Delta}$ of the phase-noise process in the SLO case must be set to $2.34^\circ$.

Fig.~\ref{fig:Diff_OutageProb_vsM_var6_log} shows the difference between the outage capacity in the CLO and the SLO cases, for a target outage probability $\varepsilon=0.1$, as a function of the number of antennas.
As expected, the gap increases as the number of antennas get large. Note that the gap does not depend on the variance of the phase-noise processes (see~\eqref{eq:PNN_miso_common_osc_memory} and~\eqref{eq:miso_multiple_osc_memory_uniform_marginal})
%

\section{Conclusions}
We studied the capacity of multiple-antenna systems affected by  phase noise. 
Specifically, we analyzed the first two terms in the high-SNR  expansion  of the capacity of  both the uplink  and the downlink channel of a system where wireless communications occur between a  base station equipped with $M$ antennas  and a single-antenna user.
Our analysis covers two different configurations: the case when the RF circuitries connected to each antenna at the base station are driven by separate local oscillators, and the case when a common oscillator drives all the antennas.  

For all four cases (uplink/downlink, common/separate oscillators) the first term in the high-SNR capacity expansion is equal to $0.5\ln(\snr)$,
whereas the second term, which we denote as phase noise number, turns out to take different values depending on which case is considered.
For the uplink channel, the  phase noise number is larger when separate oscillators are used. 
For the specific case of Wiener phase noise, a gain of at least $0.5 \ln(M)$ can be achieved. 
This gain, which is due to diversity, implies that to achieve the same throughput in the high-SNR regime, the oscillator used in the common oscillator configuration must be at least $M$ times better than any of the oscillators used in the separate configuration. 

In contrast, the phase noise number of the downlink channel is larger when a common oscillator drives all the antennas. 
This is  due to the fact that conjugate beamforming, which provides a coherent-combining gain for the common oscillator configuration, does not achieve the phase-noise number when  separate oscillators are used. 
The capacity achieving-strategy for the separate oscillator configuration turns out to be antenna selection, i.e.,
activating only the transmit antenna that yields the largest SISO high-SNR capacity, and switching off all other antennas. 
%
\begin{appendices}
\section{Proof of Theorem~\ref{thm:PNN_simo_separate_osc_memory}}
\label{sec:PNN_simo_separate_osc_memory}
As the proof consists of several steps, we organized it in three subsections. In Appendix~\ref{sec:proof_simo_multiple_osc_memoryless_capacity}, we prove Theorem~\ref{thm:PNN_simo_separate_osc_memory} for the special case of stationary memoryless  phase-noise processes $\{\theta_{m,k}\},~m=1,\dots,\MR$, with uniform marginal distribution over $[0,2\pi)$ (noncoherent system).
Building on this result, in Appendix~\ref{sec:proof_simo_multiple_osc_memoryless_arbitrary_capacity} we generalize the proof to the case of arbitrary stationary memoryless phase-noise processes (partially coherent system). For these first two cases, our bounds are tight, and the phase noise number is characterized in closed form.
Finally, we tackle the  case of phase-noise processes with memory in Section~\ref{sec:general_case_uplink_separate}.
Before we proceed further, we state in Lemma~\ref{lem:circularly_symmetric} below a property of the capacity-achieving input distribution, which will be used throughout this section.
\begin{lem}\label{lem:circularly_symmetric}
  The  process $\{x_k\}$ that achieves the capacity of the channel~\eqref{eq:simo_io} can be assumed  circularly symmetric, i.e., $\{\phase{x_k}\}$  can be taken to be a stationary memoryless process with uniform marginal distribution over $[0,2\pi)$ and independent of $\{\abs{x_k}\}$.  
\end{lem}
\begin{IEEEproof}
  The proof follows along the same lines as the proof of~\cite[Prop.~7]{moser09-06a}.
\end{IEEEproof}
\subsection{Noncoherent System}
\label{sec:proof_simo_multiple_osc_memoryless_capacity}
We focus on the case where the $M$ phase-noise processes $\{\theta_{m,k}\},m=1,\dots,\MR$ are  stationary memoryless  with uniform marginal distribution over $[0,2\pi)$. 
Stationarity and lack of memory imply that the time index $k$ in~\eqref{eq:simo_io} can be dropped and that the capacity expression in~\eqref{eq:memory_simo_capacity_definition} simplifies to 
\begin{IEEEeqnarray}{rCL}\label{eq:Memoryless_capacity_def}
  C(\snr)=\sup I(\vecy;x)
\end{IEEEeqnarray}
where
\begin{IEEEeqnarray}{rCL}\label{eq:IO-DL-SLO-noncoh}
    \vecy=\bTheta\vech x+\vecw
\end{IEEEeqnarray}
and where the supremum in~\eqref{eq:Memoryless_capacity_def} is over all  probability distributions on $x$ that satisfy 
\begin{IEEEeqnarray}{rCL}\label{eq:simo_memoryless_power_constraint}
  \Ex{}{\abs{x}^2}\leq 2\rho.
\end{IEEEeqnarray}   
We show next that the prelog and phase-noise number of the SIMO channel~\eqref{eq:IO-DL-SLO-noncoh} are 
\begin{IEEEeqnarray}{rCL}
  \eta&=&{1}/{2}, \qquad
	\chi=({1}/{2})\ln \lefto({\vecnorm{\vech}^2}/{2}\right)\label{eq:pn-DL-noncoherent}.
\end{IEEEeqnarray} 
To prove~\eqref{eq:pn-DL-noncoherent}, we establish a lower bound and an upper bound on $C(\snr)$ that match up to a $o(1)$ term as $\snr\to\infty$.
\paragraph*{Lower bound} 
To lower-bound $C(\snr)$, we evaluate the mutual information on the right-hand side (RHS) of \eqref{eq:Memoryless_capacity_def} for a specific probability distribution,
namely, we choose $\phase{x}$ to be uniformly distributed over $[0,2\pi)$, and we set $\abs{x}^2=4\rho s$ where the random variable $s$ is independent of $\phase{x}$ and is $\mathrm{Gamma}(1/2,1)$-distributed, i.e., its pdf $f_S(s)$ is given by 
\begin{IEEEeqnarray}{rCL}
  f_S(s)=e^{-s}/\sqrt{\pi s}, \quad s\geq 0.\label{eq:simo_memoryless_lower_bound_input_pdf}
\end{IEEEeqnarray}
Note that, with this choice,~\eqref{eq:simo_memoryless_power_constraint} holds with equality because $\Ex{}{s}=1/2$. 
Let $t=\vecnorm{\vecy}^2$.
The mutual information in~\eqref{eq:Memoryless_capacity_def} can be lower-bounded as follows
\begin{IEEEeqnarray}{rCL}
  I(\vecy;x)
  &\geq&I\lefto(t;x\right)=h\lefto(t\right)-h\lefto(t\given x\right).\label{eq:Memoryless_MI_LO_Entropies}
\end{IEEEeqnarray}
Here,~\eqref{eq:Memoryless_MI_LO_Entropies} follows from the data-processing inequality~\cite[Th.~2.8.1]{cover06-a}. 
Observe now that, given $x$, the random variable $t$ follows a noncentral chi-squared distribution\footnote{Normalizing the noise variance to $2$ (see Section~\ref{sec:SISO}) is crucial to obtain a chi-squared distribution. The more common unitary normalization for the noise variance would result in an additional constant factor, which is  tedious to track.} with $2\MR$ degrees of freedom and noncentrality parameter $\vecnorm{\vech x}^2$, i.e.,
\begin{IEEEeqnarray}{rCL}\label{eq:Memoryless_ttilde_EqDist}
{t}&\distas&\big|\vecnorm{\vech x}+w_1\big|^2+\sum_{m=2}^{\MR}\abs{w_m}^2
\end{IEEEeqnarray}
%
where $\distas$ denotes equality in distribution and $\{w_m\}$, $m=1,\dots,\MR$ are \iid $\jpg(0,2)$-distributed.  

Let $\phi$ be uniformly distributed over $[0,2\pi)$ and independent of $x$ and of the  $\{w_m\}$.
Using \eqref{eq:Memoryless_ttilde_EqDist}, we lower-bound the first term in \eqref{eq:Memoryless_MI_LO_Entropies} as follows: 
\begin{IEEEeqnarray}{rCL}
h\lefto(t\right)&=&h\lefto(\big|\vecnorm{\vech x}+w_1\big|^2+\sum_{m=2}^{\MR}\abs{w_m}^2\right)\label{eq:Memoryless_MI_LO_ttilde}\\
&\geq&h\lefto(\big|\vecnorm{\vech x}+w_1\big|^2\right)\label{eq:Memoryless_MI_LO_ttilde_Entr_2}\\
&=&h\lefto(e^{j\phi}\vecnorm{\vech x}+w_1\right)-\ln(\pi)\label{eq:Memoryless_MI_LO_ttilde_Entr_3}\\
&\geq&h\lefto(e^{j\phi}\vecnorm{\vech x}\right)-\ln(\pi)\\ 
&=&h\lefto(\vecnorm{\vech x}^2\right)\label{eq:Memoryless_MI_LO_ttilde_Entr_4}\\
&=&h(s)+\ln (4 \rho)+\ln\lefto(\vecnorm{\vech}^2\right)\label{eq:Memoryless_MI_LO_ttilde_Entr_4c}\\
&=&\frac{1}{2}\psi\lefto(\frac{1}{2}\right)+\frac{1}{2}+\frac{1}{2}\ln(\pi)+\ln(4 \rho)+\ln\lefto(\vecnorm{\vech}^2\right). \IEEEeqnarraynumspace\label{eq:Memoryless_MI_LO_Final_ttilde_Entr}
\end{IEEEeqnarray}
To obtain~\eqref{eq:Memoryless_MI_LO_ttilde_Entr_3},  we used first that $h(w)=h(\abs{w}^2)+\ln(\pi)$ for every circularly symmetric random variable $w$~\cite[Eq.~(320)]{lapidoth03-10a}, and then that $w_1$ is circularly symmetric, which implies that $e^{j\phi}w_1\distas w_1$; \eqref{eq:Memoryless_MI_LO_ttilde_Entr_4} follows again from~\cite[Eq.~(320)]{lapidoth03-10a}; in~\eqref{eq:Memoryless_MI_LO_ttilde_Entr_4c} we used that $h(a q)=h(q)+\ln a$ for every real-valued random variable $q$ and every positive constant $a$, and also that $\abs{x}^2=4\rho s$; finally,~\eqref{eq:Memoryless_MI_LO_Final_ttilde_Entr} holds because the differential entropy of the Gamma-distributed random variable $s$ is \cite[Eq.~(19)]{lapidoth02-10a}
\begin{IEEEeqnarray}{rCL}
h(s)&=&({1}/{2})\psi\lefto({1}/{2}\right)+{1}/{2}+({1}/{2})\ln(\pi)
\end{IEEEeqnarray}
with $\psi\lefto(\cdot\right)$ denoting Euler's digamma function.
We next upper-bound the second term of \eqref{eq:Memoryless_MI_LO_Entropies} by proceeding as in \cite{lapidoth02-10a}:
\begin{IEEEeqnarray}{rCL}
h\lefto({t}\given x\right)&\leq&
\frac{1}{2}\Ex{}{\ln\bigl(8\pi e(\MR+4\rho s\vecnorm{\vech}^2 )\bigr)}\label{eq:Memoryless_MI_LO_ttilde_Given_x_Entr_1stStep}\\
&=&\frac{1}{2}\ln(32\pi e\vecnorm{\vech}^2 \rho)+\frac{1}{2}\psi\lefto(\frac{1}{2}\right)+\landauo(1), \quad \rho\to\infty.\notag\\\label{eq:Memoryless_MI_LO_ttilde_Given_x_Entr_Final}
\end{IEEEeqnarray}
In \eqref{eq:Memoryless_MI_LO_ttilde_Given_x_Entr_1stStep} we used that the conditional variance of $t$ given $\abs{x}^2=4\snr s$ is $4(\MR+4\snr s\vecnorm{\vech}^2)$ and that the Gaussian distribution maximizes differential entropy under a
variance constraint;~\eqref{eq:Memoryless_MI_LO_ttilde_Given_x_Entr_Final} follows because  $\Ex{}{\ln s }=\psi(1/2)$ \cite[Eq.~(18)]{lapidoth02-10a}.
Substituting \eqref{eq:Memoryless_MI_LO_Final_ttilde_Entr} and \eqref{eq:Memoryless_MI_LO_ttilde_Given_x_Entr_Final} into \eqref{eq:Memoryless_MI_LO_Entropies} and then \eqref{eq:Memoryless_MI_LO_Entropies} into \eqref{eq:Memoryless_capacity_def}, we obtain
\begin{IEEEeqnarray}{rCL}\label{eq:simo_memoryless_capacity_lower_bound}
  C(\snr)&\geq& \frac{1}{2}\ln (\snr) +\frac{1}{2}\ln \lefto(\frac{\vecnorm{\vech}^2}{2}\right) +\landauo(1), \quad \snr\to\infty.
\end{IEEEeqnarray} 
%

\paragraph*{Upper Bound} 
Since the RHS of~\eqref{eq:simo_memoryless_capacity_lower_bound} coincides with the asymptotic capacity expansion for the UL-CLO case (set $h(\{\theta_k\})=\ln(2\pi)$ in~\eqref{eq:PNN_simo_common_osc_memory}), to establish~\eqref{eq:pn-DL-noncoherent} it is sufficient to show that the capacity in the UL-CLO case is no smaller than that in the UL-SLO case.
Let $\phi$ be uniformly distributed over $[0,2\pi)$ and independent of all other random variables in~\eqref{eq:IO-DL-SLO-noncoh}.
Furthermore, let 
\begin{IEEEeqnarray}{rCL}\label{eq:reduction_to_UL_CLO}
  \tilde{\vecy}=e^{j\phi}\vech x +\vecw
\end{IEEEeqnarray}
where
$\tilde{\vecy}=\tp{[\tilde{y}_1,\dots,\tilde{y}_{\MR}]}$ and all the other quantities are defined as in~\eqref{eq:IO-DL-SLO-noncoh}.
Recall also that, by definition, $\bTheta=\diag\{[e^{j\theta_1},\dots,e^{j\theta_{\MR}}]\}$.
We upper-bound the mutual information on the RHS of \eqref{eq:Memoryless_capacity_def} by proceeding as follows:
\begin{IEEEeqnarray}{rCL}\label{eq:simo_memoryless_mutual_information_upper_bound}
  I(\vecy;x)
  &\leq& I(\vecy, \{\theta_m-\phi\}_{m=1}^{\MR} ;x)\label{eq:simo_memoryless_mutual_information_upper_bound_1}\\  
  &=&I(\vecy;x\given \{\theta_m-\phi\}_{m=1}^{\MR})\label{eq:simo_memoryless_mutual_information_upper_bound_2}\\
&=&I(\tilde{\vecy};x).\label{eq:simo_memoryless_mutual_information_upper_bound_3}
\end{IEEEeqnarray}
Here,~\eqref{eq:simo_memoryless_mutual_information_upper_bound_2} follows because $x$ and $\{\theta_m-\phi\}_{m=1}^{\MR}$ are independent. 
Since~\eqref{eq:reduction_to_UL_CLO} coincides with the input-output relation for the UL-CLO case, we conclude that
\begin{IEEEeqnarray}{rCL}\label{eq:simo_memoryless_capacity_upper_bound}
  C(\snr)&\leq& \frac{1}{2}\ln (\snr) +\frac{1}{2}\ln \lefto(\frac{\vecnorm{\vech}^2}{2}\right) +\landauo(1), \quad \snr\to\infty.
\end{IEEEeqnarray}
The upper bound~\eqref{eq:simo_memoryless_capacity_upper_bound} matches the lower bound~\eqref{eq:simo_memoryless_capacity_lower_bound} up to a $o(1)$ term.
This implies~\eqref{eq:pn-DL-noncoherent}.
\subsection{Partially Coherent System}
\label{sec:proof_simo_multiple_osc_memoryless_arbitrary_capacity}
We next analyze the partially coherent case where the phase-noise processes are stationary memoryless with arbitrary marginal probability distribution (we do not require the processes to be independent across antennas), and prove that prelog and phase-noise numbers are $\eta={1}/{2}$ and
\begin{IEEEeqnarray}{rCL}
\chi&=& ({1}/{2})\ln \lefto({\vecnorm{\vech}^2}/{2}\right)+ \ln(2\pi)-h(\phi \given \phi+\vectheta)\label{eq:pn-par-cor}
\end{IEEEeqnarray}
where $\phi$ is uniformly distributed over $[0,2\pi)$ and where, by definition, $\vectheta=\tp{[\theta_1,\dots,\theta_{\MR}]}$.
Similarly to Appendix~\ref{sec:proof_simo_multiple_osc_memoryless_capacity}, we establish this result by deriving an upper bound and a lower bound on $C(\rho)=\sup I(\vecy;x)$  that  match up to a $o(1)$ term.

\paragraph*{Lower Bound} 
%
%
%
%
We choose the same input distribution as in the noncoherent case, i.e., $x$ is  circularly symmetric with $\abs{x}^2=4 \rho s$
where $s\sim \mathrm{Gamma}(1/2,1)$.
%
%
Let $\phi=\phase{x}$ and $r=\abs{x}$.
We lower-bound the mutual information in \eqref{eq:Memoryless_capacity_def} as follows
\begin{IEEEeqnarray}{rCL}
 I(\vecy;x)&=& I\lefto(\vecy;r\right)+I\lefto(\vecy;\phi\big |r\right)\\
&\geq&I\lefto(\vecnorm{\vecy}^2;r\right)+I\lefto(\vecy;\phi\big |r\right).\label{eq:simo_multiple_osc_memoryless_arbitrary_lower_MI}
\end{IEEEeqnarray}
Here, \eqref{eq:simo_multiple_osc_memoryless_arbitrary_lower_MI} holds because of the data processing inequality. 
To evaluate the first term on the RHS of~\eqref{eq:simo_multiple_osc_memoryless_arbitrary_lower_MI}, we use~\eqref{eq:simo_memoryless_capacity_lower_bound} and obtain
\begin{IEEEeqnarray}{rCL}
  I\lefto(\vecnorm{\vecy}^2;r\right)&\geq& \frac{1}{2}\ln (\snr) +\frac{1}{2}\ln \lefto(\frac{\vecnorm{\vech}^2}{2}\right) +\landauo(1), \quad \snr\to\infty.\notag\\ \label{eq:first_term_part_coh}
\end{IEEEeqnarray}
We lower-bound the second term on the RHS of~\eqref{eq:simo_multiple_osc_memoryless_arbitrary_lower_MI}  as
\begin{IEEEeqnarray}{rCL}
  \IEEEeqnarraymulticol{3}{l}{I\lefto(\vecy;\phi\big |r\right)= I\lefto(\bigl\{\abs{y_m}\bigr\}_{m=1}^{\MR},\bigl\{\phase{y_m}\bigr\}_{m=1}^{\MR};\phi\big |r\right)}\\
%
&\geq&I\lefto(\bigl\{\phase{y_m}\bigr\}_{m=1}^{\MR};\phi\big |r\right)\\
&=&I\lefto(\bigl\{\phi+\theta_m+\phase{r \abs{h_m}+w_m}\bigr\}_{m=1}^{\MR};\phi\big |r\right)\\
&=&\ln(2\pi)-h\lefto(\phi\Big|\Bigl\{\phi+\theta_m+\phase{r \abs{h_m}+w_m}\Bigr\}_{m=1}^{\MR}, r\right).\label{eq:simo_memoryless_arbitrary_lower_bound_mutual_info_lower_bound_2nd_term_lb_2} \IEEEeqnarraynumspace
\end{IEEEeqnarray}
Fix an arbitrary $\xi_0 > 0$. 
We upper-bound the second term on the RHS of \eqref{eq:simo_memoryless_arbitrary_lower_bound_mutual_info_lower_bound_2nd_term_lb_2} as follows: 
%
%
%
%
%
%
\begin{IEEEeqnarray}{rCL}
\IEEEeqnarraymulticol{3}{l}{h\lefto(\phi\Big|\Bigl\{\phi+\theta_m+\phase{r\abs{h_m}+w_m}\Bigr\}_{m=1}^{\MR}, r\right)}\\
&=&\int_0^\infty f_r(a)h\lefto(\phi\Big|\Bigl\{\phi+\theta_m+\phase{a\abs{h_m}+w_m}\Bigr\}_{m=1}^{\MR}\right)d a \label{eq:simo_memoryless_LB_Integral_1} \IEEEeqnarraynumspace\\
&=&\int_0^{\xi_0} f_r(a) h\lefto(\phi\Big|\Bigl\{\phi+\theta_m+\phase{a\abs{h_m}+w_m}\Bigr\}_{m=1}^{\MR}\right)\!d a\notag\\
&&+\int_{\xi_0}^\infty \!\!f_r(a) h\lefto(\phi\Big|\Bigl\{\phi+\theta_m+\phase{a\abs{h_m}+w_m}\Bigr\}_{m=1}^{\MR}\right)d a  \label{eq:simo_memoryless_LB_Integral_2} \IEEEeqnarraynumspace\\
&\leq& \Pr\{0 \leq r \leq \xi_0\} \notag\\ &&\times\max_{0 \leq a \leq \xi_0}  h\lefto(\phi\Big|\Bigl\{\phi+\theta_m+\phase{a\abs{h_m}+w_m}\Bigr\}_{m=1}^{\MR}\right)\notag\\
&&+\Pr\{r \geq \xi_0\}\notag\\ &&\times\max_{a \geq \xi_0}  h\lefto(\phi\Big|\Bigl\{\phi+\theta_m+\phase{a\abs{h_m}+w_m}\Bigr\}_{m=1}^{\MR}\right) \label{eq:simo_memoryless_LB_Integral_3}\\
&=& \Pr\lefto\{0 \leq s \leq \frac{\xi_0^2}{4\rho}\right\}  h\lefto(\phi\Big|\Bigl\{\phi+\theta_m+\phase{w_m}\Bigr\}_{m=1}^{\MR}\right)\notag\\
&&+\Pr\lefto\{s \geq \frac{\xi_0^2}{4\rho}\right\}h\lefto(\phi\Big|\Bigl\{\phi+\theta_m+\phase{\xi_0\abs{h_m}+w_m}\Bigr\}_{m=1}^{\MR}\right)\notag \\\label{eq:simo_memoryless_LB_Integral_4}\\
&=&h\lefto(\phi\Big|\Bigl\{\phi+\theta_m+\phase{\xi_0\abs{h_m}+w_m}\Bigr\}_{m=1}^{\MR}\right)\label{eq:simo_memoryless_LB_Integral_5}
+\landauo(1).
\end{IEEEeqnarray}
Here, in \eqref{eq:simo_memoryless_LB_Integral_4} we used that $r=\sqrt{4\rho s}$;  \eqref{eq:simo_memoryless_LB_Integral_5} holds because for every fixed $\xi_0$  
\begin{IEEEeqnarray}{rCL}
  \lim_{\snr\to\infty}\Pr\lefto\{0 \leq s \leq {\xi_0^2}/{4\rho}\right\}=0.
\end{IEEEeqnarray}
The differential entropy in~\eqref{eq:simo_memoryless_LB_Integral_5} can be made arbitrarily close to $h\lefto(\phi|\phi+\vectheta\right)$ by choosing $\xi_0$ sufficiently large.
By substituting \eqref{eq:simo_memoryless_LB_Integral_5} in \eqref{eq:simo_memoryless_arbitrary_lower_bound_mutual_info_lower_bound_2nd_term_lb_2}, \eqref{eq:first_term_part_coh} and  \eqref{eq:simo_memoryless_arbitrary_lower_bound_mutual_info_lower_bound_2nd_term_lb_2} into \eqref{eq:simo_multiple_osc_memoryless_arbitrary_lower_MI}, and by letting $\xi_0$ tend to infinity, we conclude that 
\begin{IEEEeqnarray}{rCL}\label{eq:simo_memoryless_arbitrary_capacity_lower_bound}
  C(\snr)&\geq& ({1}/{2})\ln (\snr) +({1}/{2})\ln \lefto({\vecnorm{\vech}^2}/{2}\right) + \ln(2\pi) \notag\\&&- h\lefto(\phi\big |\phi+\vectheta\right) +\landauo(1), \quad \snr\to\infty.
\end{IEEEeqnarray} 

\paragraph*{Upper bound} 
Let $r=\abs{x}$; let also $\phi=\phase{x}$, which---without loss of generality---we shall assume uniformly distributed over $[0,2\pi)$ and independent of $r$ (see Lemma~\ref{lem:circularly_symmetric}).
Using chain rule, we decompose the mutual information  $I(\vecy;x)$ as
\begin{IEEEeqnarray}{rCL}\label{eq:simo_multiple_osc_memoryless_arbitrary_upper_MI}
 I(\vecy;x)= I\lefto(\vecy;r\right)+I\lefto(\vecy;\phi\big |r\right).
\end{IEEEeqnarray}
We next upper-bound both terms on the RHS of \eqref{eq:simo_multiple_osc_memoryless_arbitrary_upper_MI}. The first term can be bounded as follows 
\begin{IEEEeqnarray}{rCL}
I\lefto(\vecy;r\right)
&\leq&I\lefto(\vecy, \bTheta ;r\right) \\
&=&I\lefto(\vecy;r\big | \bTheta\right) \label{eq:simo_multiple_osc_memoryless_arbitrary_upper_MI_1stTerm_1}\\
&=&I\lefto(\vech r e^{j\phi} + \vecw;r\right)\label{eq:simo_multiple_osc_memoryless_arbitrary_upper_MI_1stTerm_3}\\
&=&({1}/{2})\ln (\snr) +({1}/{2})\ln \lefto({\vecnorm{\vech}^2}/{2}\right) +\landauo(1). \label{eq:simo_multiple_osc_memoryless_arbitrary_upper_MI_1stTerm_4} \IEEEeqnarraynumspace
\end{IEEEeqnarray}
Here, in \eqref{eq:simo_multiple_osc_memoryless_arbitrary_upper_MI_1stTerm_1} we used that $x$ and $\vectheta$ are independent and~\eqref{eq:simo_multiple_osc_memoryless_arbitrary_upper_MI_1stTerm_4} follows because the mutual information on the RHS of~\eqref{eq:simo_multiple_osc_memoryless_arbitrary_upper_MI_1stTerm_3} coincides with that in the noncoherent UL-CLO case.\footnote{Note that $\phi$ in~\eqref{eq:simo_multiple_osc_memoryless_arbitrary_upper_MI_1stTerm_3} plays the role of uniform phase noise, although it is the phase of the transmitted signal. This is because the mutual information in~\eqref{eq:simo_multiple_osc_memoryless_arbitrary_upper_MI_1stTerm_3} is between the channel output and the amplitude of the transmitted signal.}

Next, we upper-bound the second term on the RHS of 
\eqref{eq:simo_multiple_osc_memoryless_arbitrary_upper_MI}:
\begin{IEEEeqnarray}{rCL}
I\lefto(\vecy;\phi  |r\right)&=&I(\vecy,r ; \phi)\label{eq:simo_multiple_osc_memoryless_arbitrary_upper_MI_2ndTerm_1}\\
&\leq& I\lefto( \vecy, r, \{e^{j(\theta_m+\phi)}\}_{m=1}^{\MR};\phi\right)\\
&=&I\lefto( \bigl\{e^{j(\theta_m+\phi)}\bigr\}_{m=1}^{\MR};\phi\right)\label{eq:simo_multiple_osc_memoryless_arbitrary_upper_MI_2ndTerm_2}\\
%
%
&=& \ln(2\pi) - h\lefto(\phi\big |\phi+\vectheta\right).\label{eq:simo_multiple_osc_memoryless_arbitrary_upper_MI_2ndTerm_3}
\end{IEEEeqnarray}
Here, in \eqref{eq:simo_multiple_osc_memoryless_arbitrary_upper_MI_2ndTerm_1} we used that $\phi$ and $r$ are independent and~\eqref{eq:simo_multiple_osc_memoryless_arbitrary_upper_MI_2ndTerm_2} follows because $\phi$ and the pair $(\vecy, r)$ are conditionally independent given $\{e^{j(\theta_m+\phi)}\}_{m=1}^{\MR}$.
Substituting \eqref{eq:simo_multiple_osc_memoryless_arbitrary_upper_MI_1stTerm_4} and \eqref{eq:simo_multiple_osc_memoryless_arbitrary_upper_MI_2ndTerm_3} into \eqref{eq:simo_multiple_osc_memoryless_arbitrary_upper_MI}, we obtain
\begin{IEEEeqnarray}{rCL}\label{eq:simo_memoryless_arbitrary_capacity_upper_bound}
  C(\snr)&\leq& ({1}/{2})\ln (\snr) +({1}/{2})\ln \lefto({\vecnorm{\vech}^2}/{2}\right) + \ln(2\pi) \notag\\&&- h\lefto(\phi\big |\phi+\vectheta\right) +\landauo(1), \quad \snr\to\infty.
\end{IEEEeqnarray} 

The upper bound \eqref{eq:simo_memoryless_arbitrary_capacity_upper_bound} matches the lower bound  \eqref{eq:simo_memoryless_arbitrary_capacity_lower_bound} up to a $\landauo(1)$ term. This implies~\eqref{eq:pn-par-cor}.
\subsection{Phase Noise with Memory}\label{sec:general_case_uplink_separate}
We establish~\eqref{eq:PNN_simo_separate_osc_memory} 
by proceeding similarly as in Appendix~\ref{sec:proof_simo_multiple_osc_memoryless_capacity} and~\ref{sec:proof_simo_multiple_osc_memoryless_arbitrary_capacity}.

\paragraph*{Lower Bound} 
Fix an arbitrary $\xi_0>0$ and some positive integer $\gamma$.
We evaluate the mutual information on the RHS of~\eqref{eq:memory_simo_capacity_definition} for an \iid input process $\{x_k\}$ having the same marginal distribution as in the noncoherent case, i.e., uniform phase and amplitude distributed as in~\eqref{eq:simo_memoryless_lower_bound_input_pdf}. 
Using the chain rule for mutual information and the nonnegativity of mutual information, we obtain
\begin{IEEEeqnarray}{rCL}
  I(x^n;\vecy^n)&=&\sumo{k}{n}I(x_k;\vecy^n\given x^{k-1})\\
					 &\geq&\sum_{k=\gamma+1}^{n-\gamma}I(x_k;\vecy^k\given x^{k-1}).\label{eq:memory_simo_mutual_info_lower_bound_chain_rule}
\end{IEEEeqnarray}
For every $k\in[\gamma+1,n-\gamma]$, we lower-bound $I(x_k;\vecy^k\given x^{k-1})$ as follows:
\begin{IEEEeqnarray}{rCL}
\IEEEeqnarraymulticol{3}{l}{ I\lefto(x_k;\vecy^n\given x^{k-1}\right)=I\lefto(x_k;x^{k-1},\vecy^k\right)}
 \label{eq:memory_simo_lower_bound_mutual_info_lower_bound_1}\\
&\geq&I\lefto(x_k;x_{k-\gamma}^{k-1},\vecy_{k-\gamma}^{k-1},\vecy_k\right)\label{eq:memory_simo_lower_bound_mutual_info_lower_bound_2}\\
&=&I\lefto(x_k;x_{k-\gamma}^{k-1},\vecy_{k-\gamma}^{k-1},\vectheta_{k-\gamma}^{k-1},\vecy_k\right)\notag\\
&&-\underbrace{I\lefto(x_k;\vectheta_{k-\gamma}^{k-1}\given x_{k-\gamma}^{k-1},\vecy_{k-\gamma}^{k-1},\vecy_k\right)}_{\leq \epsilon(\rho,\xi_0,\gamma)}\label{eq:memory_simo_lower_bound_mutual_info_lower_bound_3}\\
&\geq&I\lefto(x_k;x_{k-\gamma}^{k-1},\vecy_{k-\gamma}^{k-1},\vecy_k,\vectheta_{k-\gamma}^{k-1}\right)-\epsilon(\rho,\xi_0,\gamma)\\
&=&I\lefto(x_k;\vecy_k,\vectheta_{k-\gamma}^{k-1}\right)-\epsilon(\rho,\xi_0,\gamma)\label{eq:memory_simo_lower_bound_mutual_info_lower_bound_4}\\
%
%
&=&I\lefto(x_{\gamma+1};\vecy_{\gamma+1}\given\vectheta_{1}^{\gamma}\right)-\epsilon(\rho,\xi_0,\gamma). \IEEEeqnarraynumspace\label{eq:memory_simo_lower_bound_mutual_info_lower_bound_42}
\end{IEEEeqnarray}
Here, \eqref{eq:memory_simo_lower_bound_mutual_info_lower_bound_1} follows because the $\{x_k\}$ are \iid;  in \eqref{eq:memory_simo_lower_bound_mutual_info_lower_bound_3} we upper-bounded the second mutual information by a function, which we denote by $\epsilon(\snr,\xi_0,\gamma)$, that depends only on~$\snr$, $\xi_0$, and $\gamma$ and that satisfies (see Appendix~\ref{app:simo_memory_lowerbound_epsilons})
\begin{IEEEeqnarray}{rCL}
  \lim_{\xi_0 \to\infty} \lim_{\rho \to\infty} \epsilon(\rho, \xi_0,\gamma) = 0\label{eq:property_epsilon}
\end{IEEEeqnarray}
for all $\gamma$;
\eqref{eq:memory_simo_lower_bound_mutual_info_lower_bound_4} follows because $x_k$ and the pair $(\vecy_{k-\gamma}^{k-1},x_{k-\gamma}^{k-1})$ are conditionally independent given $(\vecy_k,\vectheta_{k-\gamma}^{k-1})$; finally, in \eqref{eq:memory_simo_lower_bound_mutual_info_lower_bound_42} we used stationarity and that
$x_{\gamma+1}$ and $\vectheta_{1}^{\gamma}$ are independent.
%
%
%

Substituting~\eqref{eq:memory_simo_lower_bound_mutual_info_lower_bound_42} into \eqref{eq:memory_simo_mutual_info_lower_bound_chain_rule} and then \eqref{eq:memory_simo_mutual_info_lower_bound_chain_rule} into \eqref{eq:memory_simo_capacity_definition}, 
\begin{IEEEeqnarray}{rCL}
  C(\snr)&\geq&
  I(x_{\gamma+1};\vecy_{\gamma+1} \given \vectheta_{1}^{\gamma})-\epsilon(\snr,\xi_0,\gamma).\label{eq:memory_simo_lower_bound_mutual_info_lower_bound_two_terms}
\end{IEEEeqnarray}
Apart from the side information $(\vectheta_{1}^{\gamma})$, the mutual information on the RHS of~\eqref{eq:memory_simo_lower_bound_mutual_info_lower_bound_two_terms} coincides with that of the memoryless partially coherent channel we analyzed in Appendix~\ref{sec:proof_simo_multiple_osc_memoryless_arbitrary_capacity}.
Proceeding as in Appendix~\ref{sec:proof_simo_multiple_osc_memoryless_arbitrary_capacity}, and letting $\xi_0$ tend to infinity, we obtain
\begin{IEEEeqnarray}{rCL}
  \IEEEeqnarraymulticol{3}{l}{
  I\lefto(x_{\gamma+1};\vecy_{\gamma+1}\given \vectheta_{1}^{\gamma}\right)
  \geq \frac{1}{2}\ln (\snr) +\frac{1}{2}\ln \lefto(\frac{\vecnorm{\vech}^2}{2}\right)+\ln(2\pi)}\notag\\
&&-h\lefto(\phi_{\gamma+1}\given\vectheta_{\gamma+1}+\phi_{\gamma+1},\vectheta_{1}^{\gamma}\right)+\landauo(1), \quad \snr\to\infty.\label{eq:memory_simo_lower_bound_mutual_info_lower_bound_1st_term_lower_bound}
\end{IEEEeqnarray}
Substituting~\eqref{eq:memory_simo_lower_bound_mutual_info_lower_bound_1st_term_lower_bound} into~\eqref{eq:memory_simo_lower_bound_mutual_info_lower_bound_two_terms}, using stationarity, and letting $\gamma$ tend to infinity, we finally obtain the following capacity lower bound  
\begin{IEEEeqnarray}{rCL}\label{eq:simo_memory_lowerbound_final}
  C(\snr)&\geq& ({1}/{2})\ln (\snr) +({1}/{2})\ln \lefto({\vecnorm{\vech}^2}/{2}\right) \notag\\&&+\ln(2\pi)-h\lefto(\phi_0\given\vectheta_{0}+\phi_0, ,\vectheta^{-1}_{-\infty}\right)+\landauo(1). \IEEEeqnarraynumspace
\end{IEEEeqnarray}

\paragraph*{Upper bound} We use the chain rule for mutual information on the RHS of~\eqref{eq:memory_simo_capacity_definition} and obtain
\begin{IEEEeqnarray}{rCL}\label{eq:memory_simo_mutual_info_chain_rule}
  I\lefto(x^n,\vecy^n\right)=\sumo{k}{n}I\lefto(x^n;\vecy_k\given \vecy^{k-1}\right).
\end{IEEEeqnarray}
We next proceed as in \cite[Eq.~(77)]{lapidoth06-02a} and upper-bound each term on the RHS of \eqref{eq:memory_simo_mutual_info_chain_rule} as
\begin{IEEEeqnarray}{rCL}
	I\lefto(x^n;\vecy_k\given \vecy^{k-1}\right)
&\leq&I\lefto(x^n,\vecy^{k-1};\vecy_k\right)\label{eq:simo_chain_rule_step_1}\\
&=&I\lefto(x_k, x^{k-1},\vecy^{k-1};\vecy_k\right)\label{eq:memory_simo_mutual_info_upper_bound0}\\
%
%
&\leq&I\lefto(x_k, x^{k-1},\vecy^{k-1},\vectheta^{k-1};\vecy_k\right)\\
&=&I\lefto(x_k,\vectheta^{k-1};\vecy_k\right)\label{eq:memory_simo_mutual_info_upper_bound2}\\
&=&I\lefto(x_k;\vecy_k\right)+I\lefto(\vectheta^{k-1};\vecy_k\given x_k\right)\\
%
%
&\leq& I\lefto(x_k;\vecy_k\right)+I\lefto(\vectheta^{k-1};\vecy_k, x_k,\vectheta_k\right) \IEEEeqnarraynumspace\\
&=&I\lefto(x_k;\vecy_k\right)+I\lefto(\vectheta^{k-1};\vectheta_k\right)\label{eq:memory_simo_mutual_info_upper_bound4}\\
&\leq&I\lefto(x_0;\vecy_0\right)+I\lefto(\vectheta_0;\vectheta^{-1}_{-\infty}\right)\label{eq:memory_simo_mutual_info_upper_bound5}.
\end{IEEEeqnarray}
Here, \eqref{eq:memory_simo_mutual_info_upper_bound2} holds because $\vecy_k$ and $(\vecy^{k-1},x^{k-1})$ are conditionally independent given $(x_k,\vectheta^{k-1})$; 
\eqref{eq:memory_simo_mutual_info_upper_bound4} holds because $\vectheta^{k-1}$ and $(x_k,\vecy_k)$ are conditionally independent given $\vectheta_{k}$; finally, \eqref{eq:memory_simo_mutual_info_upper_bound5} follows because of the stationarity of the phase-noise processes and the nonnegativity of mutual information. 
Substituting  \eqref{eq:memory_simo_mutual_info_upper_bound5} into \eqref{eq:memory_simo_mutual_info_chain_rule}, then \eqref{eq:memory_simo_mutual_info_chain_rule} into \eqref{eq:memory_simo_capacity_definition}, we obtain  
\begin{IEEEeqnarray}{rCL}\label{eq:simo_memory_capacity_stationary}
  C(\snr) &\leq& \sup \bigl\{I\lefto(x_0;\vecy_0\right)\bigr\}+I\lefto(\vectheta_0;\vectheta^{-1}_{-\infty}\right)\\
&\leq& ({1}/{2})\ln (\snr) +({1}/{2})\ln \lefto({\vecnorm{\vech}^2}/{2}\right) + \ln(2\pi) \notag\\&&
- h\lefto(\phi\big |\phi+\vectheta_0\right) +I\lefto(\vectheta_0;\vectheta^{-1}_{-\infty}\right) +\landauo(1)\IEEEeqnarraynumspace
\end{IEEEeqnarray}
where the last step follows from \eqref{eq:simo_memoryless_arbitrary_capacity_upper_bound}. 

\section{Proof of Theorem~\ref{thm:miso_multiple_osc_memory_upper_bound}}
\label{sec:proof_miso_multiple_osc_memory_upper_bound}
As a first step, we adapt Lemma~\ref{lem:circularly_symmetric}, which describes the structure of the capacity-achieving distribution for the SIMO phase-noise channel~\eqref{eq:simo_io} to the MISO phase-noise channel~\eqref{eq:miso_io}. 
Let $\{\phi_k\}$ be a stationary memoryless process with uniform marginal distribution over $[0,2\pi)$.
We say that a vector process $\{\vecx_k\}$ is circularly symmetric if  $\{\vecx_k e^{j\phi_k}\}\distas \{\vecx_k\}$.
\begin{lem}\label{lem:circ_symm_MISO}
  The input process that achieves the capacity of the channel~\eqref{eq:miso_io} can be assumed to be circularly symmetric.
\end{lem}
\begin{IEEEproof}
  The proof follows along the same lines as the proof of~\cite[Prop.~7]{moser09-06a}.
\end{IEEEproof}

By Lemma~\ref{lem:circ_symm_MISO}, we can express $\{\vecx_k\}$ as $\{\vecx_k=\vecnorm{\vecx_k}\hat{\vecx}_k e^{j\phi_k}\}$ where $\vecnorm{\hat{\vecx}_k}=1$ for all $k$ and $\{\phi_k\}$ is a stationary memoryless process, independent of $\{\vecnorm{\vecx_k},\hat{\vecx}_k\}$ and with uniform marginal distribution over $[0,2\pi)$.

\paragraph*{Lower Bound}
See Remark~4.
\paragraph*{Upper Bound}
We use chain rule for mutual information on the RHS of~\eqref{eq:miso_memory_capacity_def} and obtain
\begin{IEEEeqnarray}{rCL}\label{eq:chain_rule_MISO}
  I(\vecx^n;y^n)&=& \sum_{k=1}^n I(\vecx^n;y_k\given y^{k-1}).
\end{IEEEeqnarray}
Proceeding similarly to~\eqref{eq:simo_chain_rule_step_1}--\eqref{eq:memory_simo_mutual_info_upper_bound5}, we next upper-bound each term on the RHS of~\eqref{eq:chain_rule_MISO}:
\begin{IEEEeqnarray}{rCL}
  \IEEEeqnarraymulticol{3}{l}{I(\vecx^n;y_k\given y^{k-1}) \leq I(\vecx^n, y^{k-1} ;y_k)} \\
  &=& I(\vecx^k, y^{k-1} ;y_k) \label{eq:markov_MISO_1}\\
  &\leq & I(\vecx_k,\vecx^{k-1}, y^{k-1}, \vectheta^{k-1} ;y_k) \\
  &=& I(\vecx_k, \vectheta^{k-1};y_k) \label{eq:markov_MISO_2}\\
  &=& I(\vecx_k;y_k) + I(\vectheta^{k-1} ; y_k \given \vecx_k) \\
  &\leq & I(\vecx_k;y_k) + I\bigl(\vectheta^{k-1} ; y_k, \phase{\tp{\vech}{\matTheta}_k\hat{\vecx}_k} \,\big\vert \vecx_k\bigr)\\
  &=& I(\vecx_k;y_k) + I\bigl(\vectheta^{k-1} ; \phase{\tp{\vech}{\matTheta}_k\hat{\vecx}_k} \,\big\vert \vecnorm{\vecx_k}, \hat{\vecx}_k\bigr)\label{eq:markov_MISO_3}\\
  %
  %
  &\leq& I(\vecx_0;y_0) + I\bigl(\vectheta^{-1}_{-\infty} ; \phase{\tp{\vech}\matTheta_0\hat{\vecx}_0} 
  \,\big \vert 
  \vecnorm{\vecx_0}, \hat{\vecx}_0\bigr). \label{eq:stationarity_MISO} \IEEEeqnarraynumspace
\end{IEEEeqnarray}
Here,~\eqref{eq:markov_MISO_1} follows because $\vecx_{k+1}^n$ and $y_k$  are conditionally independent given $(\vecx^{k-1},y^{k-1})$;~\eqref{eq:markov_MISO_2} holds because the pair $(\vecx^{k-1},y^{k-1})$ and $y_k$  are conditionally independent given $\vectheta^{k-1}$  and $\vecx_k$; similarly,~\eqref{eq:markov_MISO_3} holds because $\vectheta^{k-1}$ and $y_k$ are conditionally independent  given $\phase{\tp{\vech}\matTheta_k\hat{\vecx}_k}$ and $\vecx_k$ and because of Lemma~\ref{lem:circ_symm_MISO}; finally, in~\eqref{eq:stationarity_MISO} we used that the phase-noise processes are stationary and that mutual information is nonnegative.

The first term on the RHS of~\eqref{eq:stationarity_MISO}, which corresponds to the mutual information achievable on a partially coherent MISO phase-noise channel, can be further upper-bounded as follows:
\begin{IEEEeqnarray}{rCL}
  \IEEEeqnarraymulticol{3}{l}{I(\vecx_0;y_0)= I(\vecnorm{\vecx_0},\hat{\vecx}_0;y_0)+I\bigl(\phi_0;y_0 \,\big\vert \vecnorm{\vecx_0},\hat{\vecx}_0\bigr)} \\
  &\leq & I(\vecnorm{\vecx_0},\hat{\vecx}_0;y_0)+I\bigl(\phi_0;y_0, \phase{\tp{\vech}\matTheta_0\vecx_0} \,\big\vert \vecnorm{\vecx_0},\hat{\vecx}_0\bigr) \\
  &=& I(\vecnorm{\vecx_0},\hat{\vecx}_0;y_0)+I\bigl(\phi_0;\phase{\tp{\vech}\matTheta_0\vecx_0} \,\big\vert \vecnorm{\vecx_0},\hat{\vecx}_0) \label{eq:markov_once_more}\\
   &=& I(\vecnorm{\vecx_0},\hat{\vecx}_0;y_0)+I\bigl(\phi_0;\phi_0+\phase{\tp{\vech}\matTheta_0\hat{\vecx}_0} \,\big\vert \vecnorm{\vecx_0}, \hat{\vecx}_0\bigr).\IEEEeqnarraynumspace\label{eq:partially_coherent_term_MISO}
\end{IEEEeqnarray}
Here,~\eqref{eq:markov_once_more} follows because $\phi_0$ and $y_0$ are conditionally independent given $\bigl(\phase{\tp{\vech}\matTheta_0\hat{\vecx}_0},\vecnorm{\vecx_0}, \hat{\vecx}_0\bigr)$.
Substituting~\eqref{eq:partially_coherent_term_MISO} into~\eqref{eq:stationarity_MISO} and using that $\phi_0$ is uniformly distributed over $[0,2\pi)$ and independent of $(\hat{\vecx}_0,\vecnorm{\vecx_0})$, we obtain
\begin{IEEEeqnarray}{rCL}
 \frac{1}{n}I(\vecx^n;y^n)&\leq& I(\vecnorm{\vecx_0},\hat{\vecx}_0;y_0) + \ln(2\pi) \notag\\&&- h(\phase{\tp{\vech}\matTheta_0\hat{\vecx}_0} \given \vecnorm{\vecx_0}, \hat{\vecx}_0, \vectheta_{-\infty}^{-1}) \\
 &\leq & \sup_{Q_{\vecnorm{\vecx_0},\hat{\vecx}_0}} \lefto\{I(\vecnorm{\vecx_0},\hat{\vecx}_0;y_0)\right\} + \ln(2\pi) \notag\\&&-\inf_{\vecnorm{\hat{\vecx}}=1} h(\phase{\tp{\vech}\matTheta_0\hat{\vecx}} \given \vectheta_{-\infty}^{-1}).\label{eq:MISO_capacity_memoryless_ub}
\end{IEEEeqnarray}
Here, the supremum is over all joint probability distributions $Q_{\vecnorm{\vecx_0},\hat{\vecx}_0}$ on $(\vecnorm{\vecx_0},\hat{\vecx}_0)$ that satisfy $\Ex{}{\vecnorm{\vecx_0}^2}\leq 2\snr$, and the infimum is over all deterministic unit-norm vectors $\hat{\vecx}$ in $\complexset^M$.

To conclude the proof, we next characterize the first term on the RHS of~\eqref{eq:MISO_capacity_memoryless_ub}.
Let 
\begin{IEEEeqnarray}{rCL}
  C_0(\snr)=\sup_{Q_{\vecnorm{\vecx_0},\hat{\vecx}_0}} I(\vecnorm{\vecx_0},\hat{\vecx}_0;y_0).
\end{IEEEeqnarray}
Then,
\begin{IEEEeqnarray}{rCL}\label{eq:prelog_memoryless_MISO}
  \lim_{\snr\to\infty} {C_0(\snr)}/{\ln(\snr)}={1}/{2}.
\end{IEEEeqnarray}
Indeed, choosing $\hat{\vecx}_0$ to be the $m$th element of the canonical basis for $\complexset^M$ (this corresponds to switching off all but the $m$th transmit antenna), the phase-noise channel 
\begin{IEEEeqnarray}{rCL}\label{eq:io_memoryless_MISO}
  y_0=e^{j\phi_0}\vecnorm{\vecx_0}\tp{\vech}\matTheta_0\hat{\vecx}_0 + w_0
\end{IEEEeqnarray}
reduces to the noncoherent SISO phase noise channel $y_0=e^{j(\phi_0+\theta_{m,0})}h_m\vecnorm{\vecx_0}+w_0$, for which the prelog is $1/2$ (see Theorem~\ref{thm:siso_capacity_Lapidoth}).
Conversely,
\begin{IEEEeqnarray}{rCL}
  C_0(\snr)&\leq& \sup_{Q_{\vecnorm{\vecx_0},\hat{\vecx}_0}} I(\vecnorm{\vecx_0},\hat{\vecx}_0;y_0,\matTheta_0) \\
  &=& \sup_{Q_{\vecnorm{\vecx_0},\hat{\vecx}_0}} I(\vecnorm{\vecx_0},\hat{\vecx}_0;y_0\given \matTheta_0) \label{eq:converse_prelog_MISO}\\
  &=& \sup_{Q_{\vecnorm{\vecx_0}}} I(\vecnorm{\vecx_0}; e^{j\phi_0}\vecnorm{\vech}\vecnorm{\vecx_0}+w_0) \label{eq:converse_prelog_MISO_1}\\
  &=& ({1}/{2})\ln (\snr) +({1}/{2})\ln \lefto({\vecnorm{\vech}^2}/{2}\right)+\landauo(1).\label{eq:converse_prelog_MISO_2}
\end{IEEEeqnarray}
Here,~\eqref{eq:converse_prelog_MISO_1} follows because setting $\hat{\vecx}_0=\herm{\matTheta}_0\conj{\vech}/\vecnorm{\vech}$ with probability 1 (w.p.1) achieves the supremum in~\eqref{eq:converse_prelog_MISO};~\eqref{eq:converse_prelog_MISO_2} follows from Theorem~\ref{thm:siso_capacity_Lapidoth}.

Fix now an arbitrary $\xi_0>0$ and let $C_{\xi_0}(\snr)$ be the capacity of the channel~\eqref{eq:io_memoryless_MISO} when the input is subject to the additional constraint that $\vecnorm{\vecx_0}\geq \xi_0$ w.p.1.
It follows from~\eqref{eq:prelog_memoryless_MISO}, from~\cite[Thm.~8]{lapidoth06-02a} and from~\cite[Thm.~4.12]{lapidoth03-10a} that
\begin{IEEEeqnarray}{rCL}\label{eq:escape-to-inf-property}
  C_0(\snr)=C_{\xi_0}(\snr)+o(1).
\end{IEEEeqnarray}
This fact is often referred to as ``escaping-to-infinity'' property of the capacity achieving distribution~\cite[Def.~4.11]{lapidoth03-10a}.

Take an arbitrary input distribution that satisfies $\Ex{}{\vecnorm{\vecx_0}^2}\leq 2\snr$ and $\vecnorm{\vecx_0}^2\geq \xi_0$ w.p.1.
Furthermore, let $q(\cdot)$ be the pdf of a $\mathrm{Gamma}(1/2,2\Ex{}{\abs{y_0}^2})$-distributed random variable:
\begin{IEEEeqnarray}{rCL}\label{eq:output_distribution_duality}
  q(r)={e^{-r/\lefto(2\Ex{}{\abs{y_0}^2}\right)}}/{\sqrt{2\pi r \Exop\bigl[\abs{y_0}^2\bigr]}}.
\end{IEEEeqnarray}
Here, the expectation is computed with respect to the probability distribution induced on $\abs{y_0}^2$ by the chosen input distribution and by the channel transition probability.
Then
\begin{IEEEeqnarray}{rCL}
  \IEEEeqnarraymulticol{3}{l}{I(\vecnorm{\vecx_0},\hat{\vecx}_0;y_0)= I(\vecnorm{\vecx_0},\hat{\vecx}_0;\abs{y_0}^2)} \label{eq:duality_MISO_1}\\
  &\leq& -\Ex{}{\ln\lefto( q(\abs{y_0}^2)\right)}-h\lefto(\abs{y_0}^2 \given \vecnorm{\vecx_0}^2,\hat{\vecx}_0\right) \label{eq:duality_MISO_2}.
\end{IEEEeqnarray}
Here,~\eqref{eq:duality_MISO_1} holds because $\abs{y_0}^2$ is a sufficient statistics for the detection of $(\hat{\vecx},\vecnorm{\vecx})$ from $y$ (recall that $\phi_0$ in~\eqref{eq:io_memoryless_MISO} is uniformly distributed over $[0,2\pi)$), and~\eqref{eq:duality_MISO_2} follows from~\cite[Thm.~5.1]{lapidoth03-10a}.
Substituting~\eqref{eq:output_distribution_duality} into~\eqref{eq:duality_MISO_2}, we obtain
\begin{multline}
  I(\vecnorm{\vecx_0},\hat{\vecx}_0;y_0)\leq {1}/{2}+({1}/{2})\ln(2\pi) +({1}/{2})\ln\lefto(\Ex{}{\abs{y_0}^2}\right)\\
  +({1}/{2})\Ex{}{\ln\lefto(\abs{y_0}^2\right)}-h\lefto(\abs{y_0}^2\given \vecnorm{\vecx_0},\hat{\vecx}_0\right).\label{eq:duality_memoryless_MISO}
\end{multline}
Note now that
\begin{IEEEeqnarray}{rCL}
  \Ex{}{\abs{y_0}^2}&\leq& 2\snr \Ex{}{\abs{\tp{\vech}\matTheta_0\hat{\vecx}_0}^2}+2 \\
  &\leq & 2\snr \sup_{\hat{\vecx}} \Ex{}{\abs{\tp{\vech}\matTheta_0\hat{\vecx}}^2}+2 \label{eq:ub_energy_MISO}
\end{IEEEeqnarray}
where the supremum is over all deterministic unit-norm vectors $\hat{\vecx}$ in $\complexset^M$.
Furthermore,
\begin{IEEEeqnarray}{rCL}
  \IEEEeqnarraymulticol{3}{l}{({1}/{2})\Ex{}{\ln\lefto(\abs{y_0}^2\right)}-h(\abs{y_0}^2\given \vecnorm{\vecx_0},\hat{\vecx}_0)}\\
  &\leq&({1}/{2})\Ex{}{\ln\lefto(\abs{y_0}^2\right)}-h(\abs{y_0}^2\given \vecnorm{\vecx_0},\hat{\vecx}_0,\matTheta_0) \\
 &\leq & \sup_{\hat{\vecx}}\sup_{\xi\geq\xi_0} \Big\{({1}/{2})\Ex{}{\ln\lefto(\abs{\xi\tp{\vech}\matTheta_0\hat{\vecx}+w}^2\right)}\notag\\&&\quad\quad\quad\quad\quad-h\lefto(\abs{\xi\tp{\vech}\matTheta_0\hat{\vecx}+w}^2 \,\big\vert\, \matTheta_0\right)\Big\}.\label{eq:term_escape_to_infty}
\end{IEEEeqnarray}

Since for every fixed $\hat{\vecx}$,~\cite[Eq.~(9)]{lapidoth02-10a}
\begin{IEEEeqnarray}{rCL}
  \IEEEeqnarraymulticol{3}{l}{
  \lim_{\xi\to\infty} \frac{1}{2}\Ex{}{\ln\lefto(\abs{\xi\tp{\vech}\matTheta_0\hat{\vecx}+w}^2\right)}-h\lefto(\abs{\xi\tp{\vech}\matTheta_0\hat{\vecx}+w}^2\,\big\vert\, \matTheta_0\right)} \nonumber \\ \quad=-({1}/{2})\ln(8\pi e)\label{eq:escape-to-inf-limit}
\end{IEEEeqnarray}
we can make~\eqref{eq:term_escape_to_infty} arbitrarily close to $-0.5 \ln(8\pi e)$ by choosing $\xi_0$ sufficiently large.

Substituting~\eqref{eq:ub_energy_MISO} and~\eqref{eq:term_escape_to_infty} into~\eqref{eq:duality_memoryless_MISO}, then using~\eqref{eq:escape-to-inf-limit} and~\eqref{eq:escape-to-inf-property}, and finally letting $\xi_0$ tend to infinity,  we obtain
\begin{IEEEeqnarray}{rCL}
  C_0(\snr)\leq \frac{1}{2}\ln (\snr) +\sup_{\hat{\vecx}}\frac{1}{2}\ln\lefto({\Ex{}{\abs{\tp{\vech}\matTheta\hat{\vecx}}^2}}/{2}
  \right) +\landauo(1).\label{eq:asymptotic-memoryless-MISO} \IEEEeqnarraynumspace
\end{IEEEeqnarray}
Substituting~\eqref{eq:asymptotic-memoryless-MISO} into~\eqref{eq:MISO_capacity_memoryless_ub}, we obtain the desired result.

\section{Proof of Theorem~\ref{thm:miso_multiple_osc_memory_uniform_marginal}}
\label{sec:proof_miso_multiple_osc_memory_uniform_marginal}
%
We start by evaluating the second term on the RHS of~\eqref{eq:miso_PNN_upper_bound_UB}. 
Let $\hat{\vecx}=\tp{[\hat{x}_1,\dots,\hat{x}_M]}$.
Then
\begin{IEEEeqnarray}{rCL}
  \sup_{\vecnorm{\hat{\vecx}}=1}\Ex{}{\abs{\tp{\vech}\bTheta_0 \hat{\vecx}}^2} &=& 
  \sup_{\vecnorm{\hat{\vecx}}=1} \Ex{}{\Bigl\lvert \sum_{m=1}^M  e^{j\theta_{m,0}}h_m\hat{x}_m\Bigr\rvert^2} \IEEEeqnarraynumspace\\
  &=&\underset{\vecnorm{\hat{\vecx}}=1}{\sup }\sumo{m}{\MT} \abs{h_{m}}^2 \abs{\hat{x}_{m}}^2\label{eq:zero-mean-indep-assumption}\\
  &=&\underset{m=1,\dots,\MT}{\max}\abs{h_m}^2.\label{eq:second_term_MISO_SLO_unif}
\end{IEEEeqnarray}
In~\eqref{eq:zero-mean-indep-assumption} we used that the random variables $\{\theta_{m,0}\}$, $m=1,\dots,M$ are independent and uniformly distributed over $[0,2\pi)$, which implies that $\Ex{}{e^{j\theta_m}}=0$, $m=1,\dots,M$.

We next bound the third term on the RHS of~\eqref{eq:miso_PNN_upper_bound_UB} as follows:
\begin{IEEEeqnarray}{rCL}
  \underset{\vecnorm{\hat{\vecx}}=1}{\inf}\Bigl\{h\lefto(\phase{\tp{\vech}\bTheta_0 \hat{\vecx}} \given \vectheta_{-\infty}^{-1}\right)\Bigr\}\notag \\
  &&\hspace{-3cm}\geq\inf_{\vecnorm{\hat{\vecx}}=1}h\lefto(\phase{\phxhato} \big | \vectheta_{-\infty}^{-1}, \theta_{2,0},\dots,\theta_{M,0} \right)\IEEEeqnarraynumspace\\
  &&\hspace{-3cm}=h\lefto(\theta_{1,0}\given  \vectheta_{-\infty}^{-1}\right)\label{eq:miso_multiple_osc_memory_upper_bound_uniform_marginal_entropy_bound_2}\\    &&\hspace{-3cm}=h\lefto(\{\theta_k\}\right).\label{eq:miso_multiple_osc_memory_upper_bound_uniform_marginal_entropy_bound}
\end{IEEEeqnarray}
Here in~\eqref{eq:miso_multiple_osc_memory_upper_bound_uniform_marginal_entropy_bound_2} we used that the phase-noise processes are independent, and in~\eqref{eq:miso_multiple_osc_memory_upper_bound_uniform_marginal_entropy_bound} that they are identically distributed.
Substituting \eqref{eq:second_term_MISO_SLO_unif} and \eqref{eq:miso_multiple_osc_memory_upper_bound_uniform_marginal_entropy_bound} in \eqref{eq:miso_PNN_upper_bound_UB}, we obtain the following upper bound on the phase-noise number 
\begin{IEEEeqnarray}{rCL}
  \chi_\DLSEP&\leq&\frac{1}{2}\ln\lefto( {\underset{m=1,\dots,\MT}{\max}\abs{h_m}^2}/{2}\right)+\ln(2\pi)-h\lefto(\{\theta_k\}\right).\IEEEeqnarraynumspace\label{eq:miso_multiple_osc_memory_upper_bound_uniform_marginal_final}
\end{IEEEeqnarray}
This bound is tight.
Indeed consider the transmission scheme where only the transmit antenna that experiences the largest channel gain $\underset{m=1,\dots,\MT}{\max}\abs{h_m}^2$ is switched on, whereas all other antennas are switched off.
By Theorem~\ref{thm:siso_capacity_Lapidoth}, the phase-noise number achievable with this transmission scheme coincides with the RHS of~\eqref{eq:miso_multiple_osc_memory_upper_bound_uniform_marginal_final}. %
This concludes the proof of Theorem~\ref{thm:miso_multiple_osc_memory_uniform_marginal}.
%
\section{}\label{app:simo_memory_lowerbound_epsilons}
We shall prove that  
\begin{IEEEeqnarray}{rCL}\label{eq:eps_1}
I\lefto(x_k;\vectheta_{k-\gamma}^{k-1}\given x_{k-\gamma}^{k-1},\vecy_{k-\gamma}^{k-1},\vecy_k,\vecy_{k+1}^{k+\gamma}\right)\leq \epsilon(\rho,\xi_0,\gamma)
\end{IEEEeqnarray}
where $\epsilon(\rho,\xi_0,\gamma)$ is a function that depend only on $\rho$,~$\xi_0$, and $\gamma$ and that satisfies~\eqref{eq:property_epsilon}
for all $\gamma$.

We proceed as follows:
\begin{IEEEeqnarray}{rCL}
\IEEEeqnarraymulticol{3}{l}{I\lefto(x_k;\vectheta_{k-\gamma}^{k-1}\given x_{k-\gamma}^{k-1},\vecy_{k-\gamma}^{k-1},\vecy_k,\vecy_{k+1}^{k+\gamma}\right)} \notag\\
&=&h\lefto(\vectheta_{k-\gamma}^{k-1}\given x_{k-\gamma}^{k-1},\vecy_{k-\gamma}^{k-1},\vecy_k,\vecy_{k+1}^{k+\gamma}\right)\notag\\
&&\!\!-h\lefto(\vectheta_{k-\gamma}^{k-1}\given x_k, x_{k-\gamma}^{k-1},\vecy_{k-\gamma}^{k-1},\vecy_k,\vecy_{k+1}^{k+\gamma}\right)\\
&\leq& h\lefto(\vectheta_{k-\gamma}^{k-1}\given  x_{k-\gamma}^{k-1},\vecy_{k-\gamma}^{k-1}\right)\notag\\
&&\!\!-h\lefto(\vectheta_{k-\gamma}^{k-1}\given x_k, x_{k-\gamma}^{k-1},\vecy_{k-\gamma}^{k-1},\vecy_k,\vecy_{k+1}^{k+\gamma},x_{k+1}^{k+\gamma},\vectheta_{k}^{k+\gamma}\right) \IEEEeqnarraynumspace\\
&=& h\lefto(\vectheta_{k-\gamma}^{k-1}\given  x_{k-\gamma}^{k-1},\vecy_{k-\gamma}^{k-1}\right)\notag\\
&&\!\!-h\lefto(\vectheta_{k-\gamma}^{k-1}\given  x_{k-\gamma}^{k-1},\vecy_{k-\gamma}^{k-1},\vectheta_{k}^{k+\gamma}\right)\\
&=&I\lefto(\vectheta_{k-\gamma}^{k-1};\vectheta_{k}^{k+\gamma}\given  x_{k-\gamma}^{k-1},\vecy_{k-\gamma}^{k-1}\right)\\
&=&h\lefto(\vectheta_{k}^{k+\gamma}\given  x_{k-\gamma}^{k-1},\vecy_{k-\gamma}^{k-1}\right)\notag \\
&&\!\!-h\lefto(\vectheta_{k}^{k+\gamma}\given  x_{k-\gamma}^{k-1},\vecy_{k-\gamma}^{k-1},\vectheta_{k-\gamma}^{k-1}\right)\\
&=&h\lefto(\vectheta_{k}^{k+\gamma}\given  x_{k-\gamma}^{k-1},\bigl\{\phase{\bTheta_l\vech x_l+\vecw_l}\bigr\}_{l=k-\gamma}^{k-1}\right)\notag\\
&&\!\!-h\lefto(\vectheta_{k}^{k+\gamma}\given \vectheta_{k-\gamma}^{k-1}\right)\\
%
%
&\leq& h\lefto(\vectheta_{k}^{k+\gamma}\given \{\phase{\bTheta_l\vech \xi_0+\vecw_l}\}_{l=k-\gamma}^{k-1}\right)\notag\\
&&\!\!-h\lefto(\vectheta_{k}^{k+\gamma}\given \vectheta_{k-\gamma}^{k-1}\right)+o(1),\quad \snr \to \infty.
\label{eq:memory_simo_lower_bound_epsilon_1}
\end{IEEEeqnarray}
Here, \eqref{eq:memory_simo_lower_bound_epsilon_1} follows by a multivariate extension of the steps taken in \eqref{eq:simo_memoryless_LB_Integral_1}--\eqref{eq:simo_memoryless_LB_Integral_5}. The RHS of \eqref{eq:memory_simo_lower_bound_epsilon_1} depends only on $\gamma$ and $\xi_0$ and can be made arbitrarily close to zero by choosing $\xi_0$ sufficiently large.

\end{appendices}
\bibliographystyle{IEEEtran}
\bibliography{IEEEabrv,publishers,confs-jrnls,reference}
\begin{IEEEbiography}[{\includegraphics[width=1in,height=1.25in,clip,keepaspectratio]{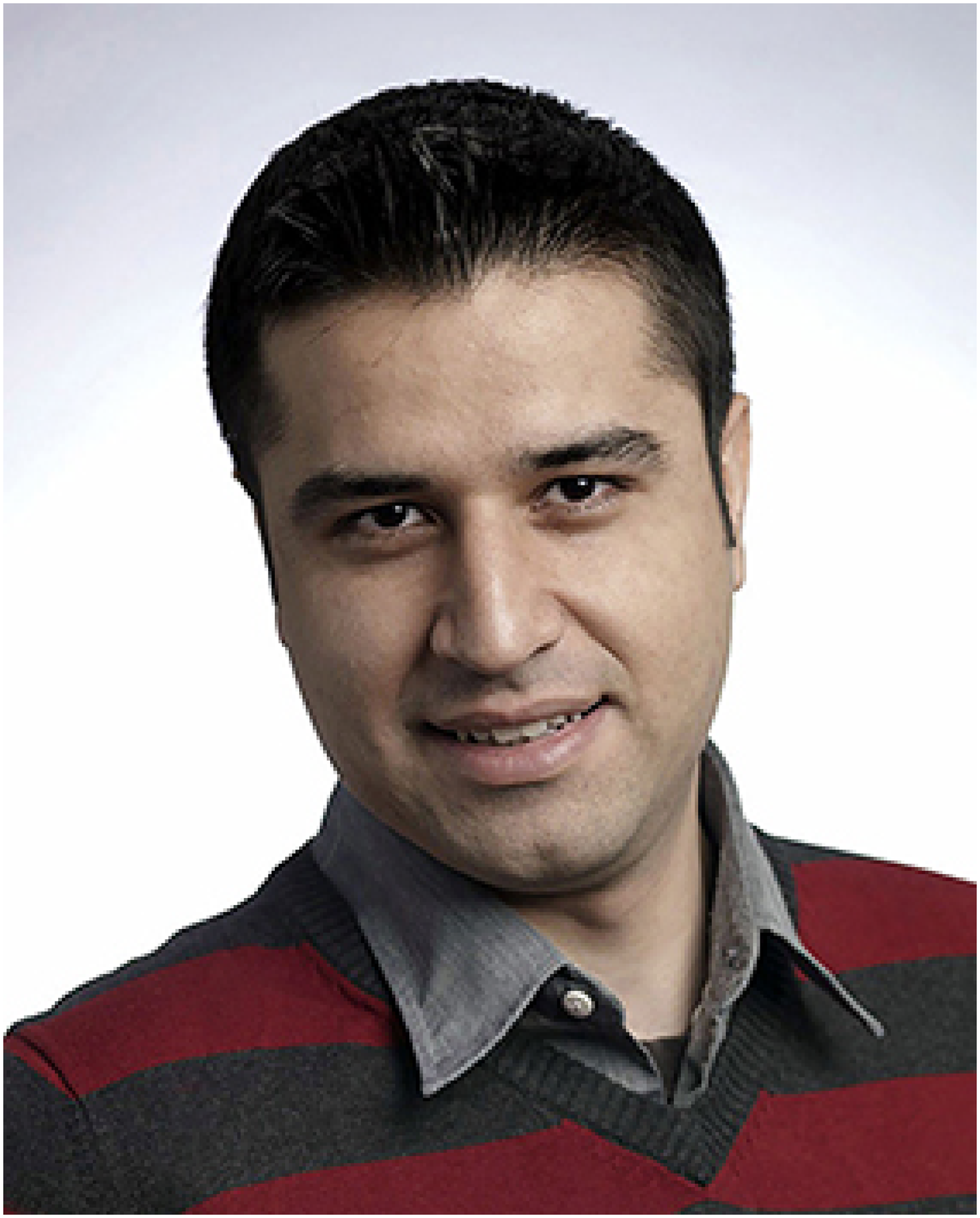}}]{M. Reza Khanzadi}(S'10)
received his M.Sc. degree in communication engineering from Chalmers University of Technology, Gothenburg, Sweden, in 2010. He is currently a Ph.D. candidate at the Department of Signals and Systems in collaboration with the Department of Microtechnology and Nanoscience of the same university. From October to December 2014, he was a Research Visitor in the University of Southern California, Los Angeles, CA. 
Bayesian inference, statistical signal processing, and information theory are his current research interests. His Ph.D. project is mainly focused on radio frequency oscillator modeling, oscillator phase noise estimation/compensation, and determining the effect of oscillator phase noise on the performance of communication systems. 
He has been the recipient of the S2 Pedagogical Prize 2012 from Department of Signals and Systems, as well as a 2013 and 2014 Ericsson's Research Foundation grants, 2014 Chalmers Foundation grant, and 2014 Gothenburg Royal Society of Arts and Sciences grant. 
\end{IEEEbiography}

\begin{IEEEbiography}[{\includegraphics[width=1in,height=1.25in,clip,keepaspectratio]{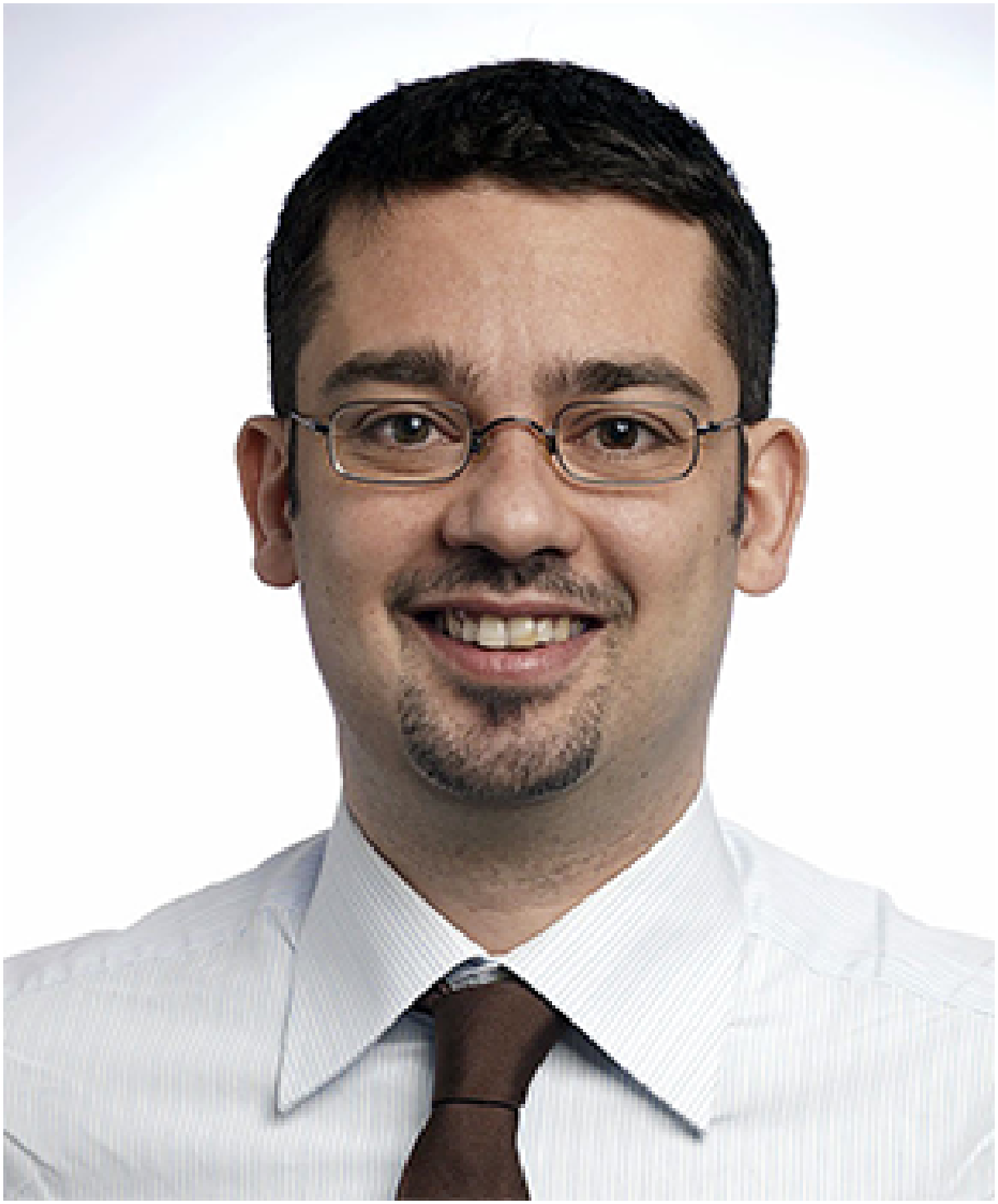}}]{Giuseppe Durisi}(S'02, M'06, SM'12)
received the Laurea degree summa cum laude and the Doctor degree both from Politecnico di Torino, Italy, in 2001 and 2006, respectively. From 2002 to 2006, he was with Istituto Superiore Mario Boella, Torino, Italy. From 2006 to 2010 he was a postdoctoral researcher at ETH Zurich, Zurich, Switzerland. Since 2010, he has been with Chalmers University of Technology, Gothenburg, Sweden, where is now associate professor. He is also guest researcher at Ericsson, Sweden. 

Dr. Durisi is a senior member of the IEEE. He is the recipient of the 2013 IEEE ComSoc Best Young Researcher Award for the Europe, Middle East, and Africa Region, and is co-author of a paper that won a ``student paper award'' at the 2012 International Symposium on Information Theory, and of a paper that won the 2013 IEEE Sweden VT-COM-IT joint chapter best student conference paper award.  From 2011 to 2014 he served as publications editor for the IEEE Transactions on Information Theory. His research interests are in the areas of communication and information theory, and compressed sensing. 

\end{IEEEbiography}

\begin{IEEEbiography}[{\includegraphics[width=1in,height=1.25in,clip,keepaspectratio]{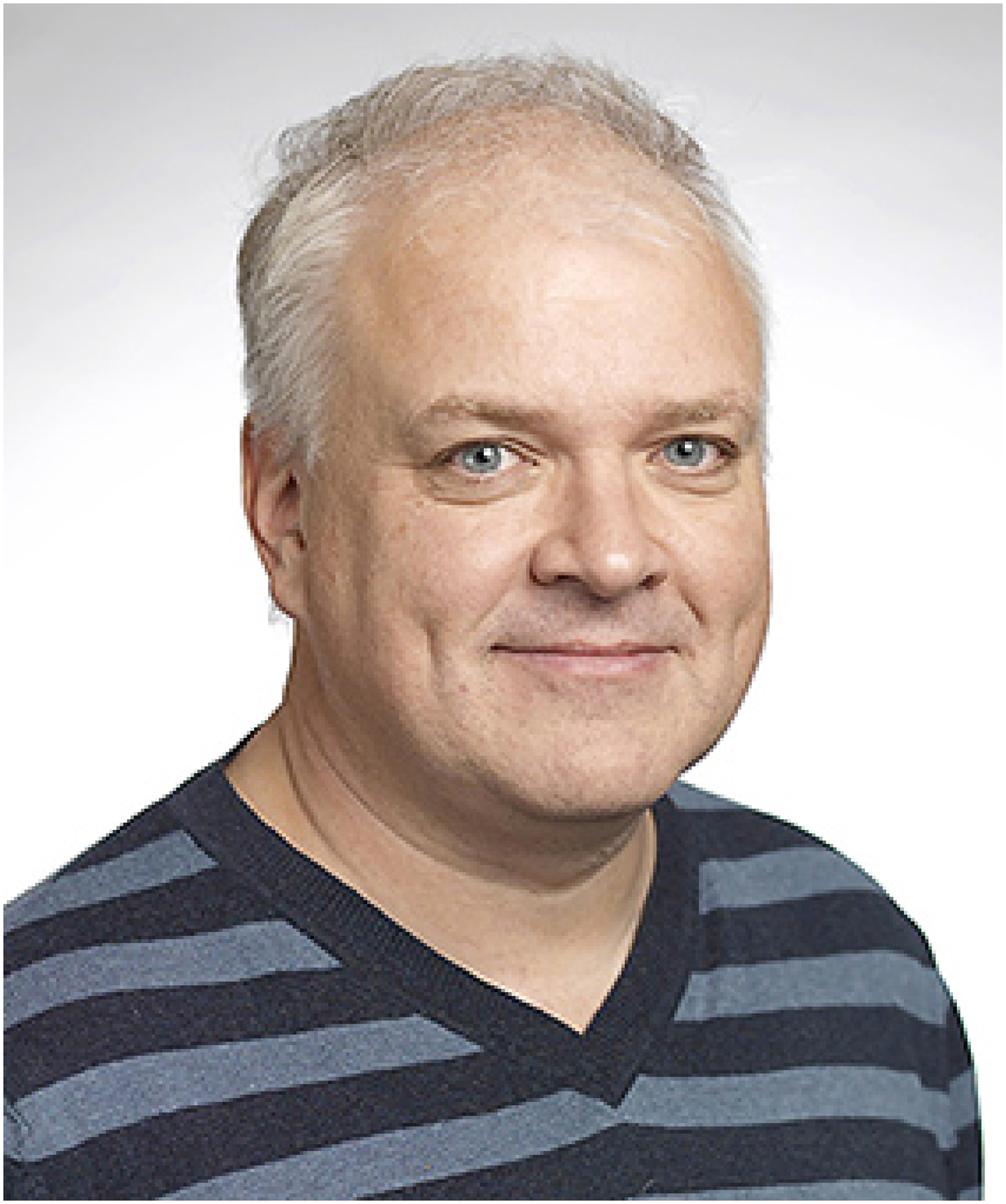}}]{Thomas Eriksson}
Thomas Eriksson received the Ph.D. degree in Information Theory in 1996, from Chalmers University of Technology, Gothenburg, Sweden. From 1990 to 1996, he was at Chalmers. In 1997 and 1998, he was at AT\&T Labs - Research in Murray Hill, NJ, USA, and in 1998 and 1999 he was at Ericsson Radio Systems AB, Kista, Sweden. Since 1999, he has been at Chalmers University, where he is a professor in communication systems. Further, he was a guest professor at Yonsei University, S. Korea, in 2003-2004. He is currently vice head of the department of Signals and Systems at Chalmers, with responsibility for undergraduate and master education. His research interests include communication, data compression, and modeling and compensation of non-ideal hardware components (e.g. amplifiers, oscillators, modulators in communication transmitters and receivers).
\end{IEEEbiography}

\end{document}